\documentclass{aa}
\usepackage{graphicx}
\usepackage[varg]{txfonts}

\usepackage{natbib}
\bibpunct{(}{)}{;}{a}{}{,}

\usepackage{color}
\usepackage{mathrsfs}
\usepackage{amsbsy}
\usepackage{mathtools}
\usepackage{tabularx}
\usepackage{multirow}
\definecolor{comm}{rgb}{0,0.7,0}

\definecolor{old}{rgb}{0.6,0.4,0.4}

\definecolor{new}{rgb}{0.5,0.2,0.5}

\definecolor{done}{rgb}{1,0.5,0}

\definecolor{todo}{rgb}{0.8,0,0}

\definecolor{idea}{rgb}{0.3,0.5,1}

\allowdisplaybreaks

\begin{document} 

\title{Exploring extreme brightness variations in blue supergiant MACHO\,80.7443.1718: Evidence for companion-driven enhanced mass loss\thanks{Based on observations made with the High-Resolution Spectrograph (HRS) at Southern African Large Telescope (SALT) and the Las Cumbres Observatory Global Telescope (LCOGT) network.}}

\author{P. A. Ko{\l}aczek-Szyma\'nski\inst{1,2}\fnmsep\thanks{This research was supported by the University of Liege under Special Funds for Research, IPD-STEMA Programme.}
\and P. {\L}ojko\inst{1}
\and A. Pigulski\inst{1}
\and T. R\'{o}\.{z}a\'nski\inst{1,3}
\and D. Mo\'{z}dzierski\inst{1}
}
\institute{University of Wroc\l aw, Faculty of Physics and Astronomy, Astronomical Institute, ul. Kopernika 11, 51-622 Wroc\l aw, Poland
\email{piotr.kolaczek-szymanski@uwr.edu.pl}
\and
University of Li\`ege (ULi\`ege), Space sciences, Technologies and Astrophysics Research (STAR), Groupe d'astrophysique des hautes \'energies (GAPHE), Quartier Agora (B5c, Institut d’Astrophysique et de G\'eophysique), All\'ee du 6 Ao\^ut
19c, B-4000 Sart Tilman, Li\`ege, Belgium
\and
Australian National University, Research School of Astronomy\& Astrophysics, Cotter Rd., Weston, ACT 2611, Australia
}
  \date{Received 28.09.2023; Accepted 05.03.2024}
\abstract
{Evolution of massive stars is dominated by interactions within binary and multiple systems. In order to accurately model this evolution, it is necessary to investigate all possible forms of interaction in binary systems that may affect the evolution of the components. One of `laboratories' plausible for this kind of investigation is the massive eccentric binary system MACHO\,80.7443.1718 (ExtEV), which exhibits an exceptionally large amplitude of light variability close to the periastron passage of its 32.8-day orbit.}
{We examine whether the light variability of the ExtEV can be explained by a wind-wind collision (WWC) binary system model. We also critically review other models proposed to explain the light curve of ExtEV.} 
{We conducted an analysis of (i) broadband multi-color photometry of ExtEV spanning a wide range of wavelengths from ultraviolet to near-infrared, (ii) time-series space photometry from Transiting Exoplanet Survey Satellite (TESS), (iii) ground-based Johnson $UBV$ photometry, and (iv) time-series high-resolution spectroscopy. To derive the parameters of the primary component of the system, we fitted spectral energy distribution (SED) and calculated evolutionary models of massive stars that included mass loss. Using radial-velocity data, we determined the spectroscopic parameters of the system. We also fitted an analytical model of light variations to the TESS light curve of ExtEV.}
{The ExtEV system exhibits an infrared excess, indicating an increased mass-loss rate. The system does not match the characteristics of B[e] stars, however. We rule out the possibility of the presence of a Keplerian disk around the primary component. We also argue that the scenario with periodic Roche-lobe overflow at periastron may not be consistent with the observations of ExtEV. Analysis of SED suggests that the primary component has a radius of about 30\,R$_\sun$ and a luminosity of $\sim6.6\cdot 10^5$\,L$_\odot$. With the analysis of the radial-velocity data, we refine the orbital parameters of ExtEV and find evidence for the presence of a tertiary component in the system. Using evolutionary models we demonstrate that the primary component's mass is between 25 and 45\,M$_\sun$. We successfully reproduce the light curve of ExtEV with our analytical model, showing that the dominant processes shaping its light curve can be attributed to atmospheric eclipse and light scattered in the WWC cone. We also estimate the primary's mass-loss rate due to stellar wind for $4.5\cdot 10^{-5}$\,M$_\sun\,{\rm yr}^{-1}$.}
{ExtEV is most likely not an extreme eccentric ellipsoidal variable, but rather an exceptional WWC binary system. The mass loss rate we derived exceeds theoretical predictions by up to two orders of magnitude. This implies that the wind in the system is likely enhanced by tidal interactions, rotation, and possibly also tidally excited oscillations. Therefore, ExtEV represents a rare evolutionary phase of a binary system that may help to understand the role of a companion-driven enhanced mass loss in the evolution of massive binary systems.}
\titlerunning{Extreme wind mass loss in MACHO\,80.7443.1718}
\authorrunning{Ko{\l}aczek-Szyma\'nski et al.}
\maketitle

\section{Introduction}\label{sect:introduction}
There is a general agreement that most massive stars with initial masses greater than  $\sim$8\,M$_\sun$ are in binary and multiple systems \citep{2013ARA&A..51..269D,2014ApJS..215...15S,2017ApJS..230...15M,2017A&A...598A..84A}. Their evolution is therefore inextricably linked to their binarity \citep{2012Sci...337..444S,2013ApJ...764..166D,2022ARA&A..60..455E}. The presence of a relatively close companion not only enables mass transfer at various stages of the evolution of a massive star \citep[e.g.][and references therein]{2022A&A...659A..98S}, but can also affect the rate of mass loss due to line-driven stellar wind \citep[e.g.][]{2020ApJ...902...85M,2022ARA&A..60..203V} or induce high-amplitude tidally excited oscillations \citep[TEOs,][]{1995ApJ...449..294K,2017MNRAS.472.1538F,2021A&A...647A..12K,2022ApJ...928..135W}. In particular, the intensity of stellar wind has important implications for the evolutionary track of a massive star \citep[e.g.][]{2014ARA&A..52..487S}. In this context, the recent findings presented by \cite{2023ApJ...948..111F} seem particularly important. The authors indicate that massive binary systems with increased mass loss rate are more efficient producers of potassium and other elements of similar atomic numbers than the evolutionary channels of single massive stars. Unfortunately, there are still many uncertainties regarding theoretical predictions of mass loss rates in massive stars due to their clumped, radiation-driven winds, especially when they leave the main sequence and become blue supergiants \citep[SGs; e.g.][]{2001A&A...369..574V,2021A&A...647A..28K,2023A&A...676A.109B,2023arXiv230313058G}. Therefore, any opportunity to observe massive binary systems during the ongoing process of envelope stripping due to mass transfer or stellar wind enhanced by the presence of a companion is particularly valuable for verifying theoretical studies. In this context, MACHO\,80.7443.1718 is of particular interest.

MACHO\,80.7443.1718 ($V\approx$ 13.6\,mag) is an eccentric (eccentricity $e\approx 0.5$) single-lined spectroscopic binary system with an orbital period of 32.83\,d \citep[][hereafter J21]{2021MNRAS.506.4083J}, located in the Large Magellanic Cloud (LMC). Discovered as variable in the MAssive Compact Halo Object survey \citep[MACHO,][]{1999PASP..111.1539A,2001yCat.2247....0M} was classified as an eclipsing binary. It was later reclassified by \citet[][hereafter J19]{2019MNRAS.489.4705J} as an eccentric ellipsoidal variable (EEV) through the analysis of All-Sky Automated Survey for SuperNovae \citep[ASAS-SN,][]{2017PASP..129j4502K} data. The primary component of this system is a B0\,Iae-type SG with an effective temperature of 30\,000\,K. Both the position of the primary component in the Hertzsprung-Russell (HR) diagram (J21) and the properties of its photometric stochastic va\-ria\-bi\-li\-ty \citep[][hereafter KS22]{2022A&A...659A..47K} suggest that the star has already evolved off the main sequence (MS) and entered the Hertzsprung gap. Based on the analysis of time-series spectra, the identification of strong emission in hydrogen Balmer lines, and the presence of forbidden iron and oxygen emission lines, J21 also argued that the primary component of MACHO\,80.7443.1718 could be B[e] SG \citep{1998A&A...340..117L,2019Galax...7...83K}.

Despite attempts, J21 failed to detect the lines of the secondary component in the composite spectrum. The invisible companion is most likely an O-type dwarf. The system is unusual for at least two reasons. First, the total range of brightness variations in the system is about 0.4\,mag, which is extreme for EEVs with MS or post-MS components (J19, their Fig.~2 and \cite{2022ApJ...928..135W}, their Fig.~11). For this reason, J19 dubbed it an `extreme' heartbeat star -- a term occasionally used to describe EEVs \citep[e.g.][]{2016ApJ...829...34S}. Neither J21 nor KS22 was able to explain such a large range of brightness variations with proximity effects using the standard models implemented in the PHysics Of Eclipsing BinariEs \citep[PHOEBE,][]{2016ApJS..227...29P,2018ApJS..237...26H,2020ApJS..247...63J,2020ApJS..250...34C} code. All these attempts yielded amplitudes at least an order of magnitude smaller than those actually observed in MACHO\,80.7443.1718. Secondly, the system shows the presence of multiple TEOs with relatively high amplitudes (J19, J21, KS22). Furthermore, MACHO\,80.7443.1718 is the first and so far the only EEV in which significant and relatively abrupt changes in both TEOs amplitudes and frequencies have been discovered (KS22). For simplicity, from this point on we refer to MACHO\,80.7443.1718 as ExtEV (meaning `extreme eccentric variable').

The true nature of the observed `extreme' brightness changes in ExtEV remains a mystery. J21 has suggested that the `heartbeat effect' seen in its light curve may be related to the presence of a circumstellar disk around the primary component. More recently, \citet[][herafter ML23]{MacLeod23} explored the possibility of nonlinear tidal wave breaking \citep{2022ApJ...937...37M} on the surface of the primary component as the source of the system's unique brightness variations. Using hydrodynamical modeling, the authors were able to obtain close-to-real amplitude only by assuming a very rapid rotation rate of the primary component.

The present study was inspired by the striking similarity in the shape of the light curve of ExtEV and the light curve of HD\,38282, a massive eccentric binary system consisting of two Wolf-Rayet (WR) stars with a total mass of about 140\,M$_\sun$ and located in the Tarantula Nebula \citep{2013MNRAS.432L..26S,2021A&A...650A.147S}. \cite{2021A&A...650A.147S} has shown that very dense stellar winds outflowing from both components of HD\,38282 cause atmospheric eclipses when the light emitted by the more distant component is attenuated by the wind of the star that is closer to the observer \citep{1994ApJ...436..859A,1996AJ....112.2227L,2005ApJ...628..953S}. In addition, the colliding winds form a turbulent and shocked wind-wind collision (WWC) cone \citep{2011A&A...530A.119P,2012ApJ...746...73T,2021Galax..10....4K}, which may be responsible for the excess emission near the periastron passage. These combined effects can result in a light curve that mimics real EEVs, even those that undergo nonlinear tidal wave breaking on their surface. We investigate in our paper whether similar phenomena may be at work in the much less massive ExtEV system, whose total mass lies between $\sim$35\,M$_\sun$ and $\sim$60\,M$_\sun$, and explain the origin of this system's unique light curve.

The paper is organized as follows. In Sect.~\ref{sect:data}, we describe the origin and preparation of the photometric and spectroscopic data used in our study. In the next section, we analyze the spectral energy distribution (SED) of the ExtEV. In Sect.~\ref{sect:results}, we discuss various properties of the ExtEV, including its potential status as a B[e] SG, the initial and current mass of the primary component, and the orbital parameters. Based on the information from the previous section, we present in Sect.~\ref{sect:lc-model} the modeling of the ExtEV light curve using an analytical model that accounts for phenomena typical of WWC binary systems. We discuss selected aspects of our results in Sect.~\ref{sect:discussion}, while a summary and final conclusions are provided in Sect.~\ref{sect:summary-and-conclusions}.

\section{Observations}\label{sect:data}

\subsection{Time-series photometry}\label{sect:time-series photometry}
\begin{figure*}
   \centering
   \includegraphics[width=\hsize]{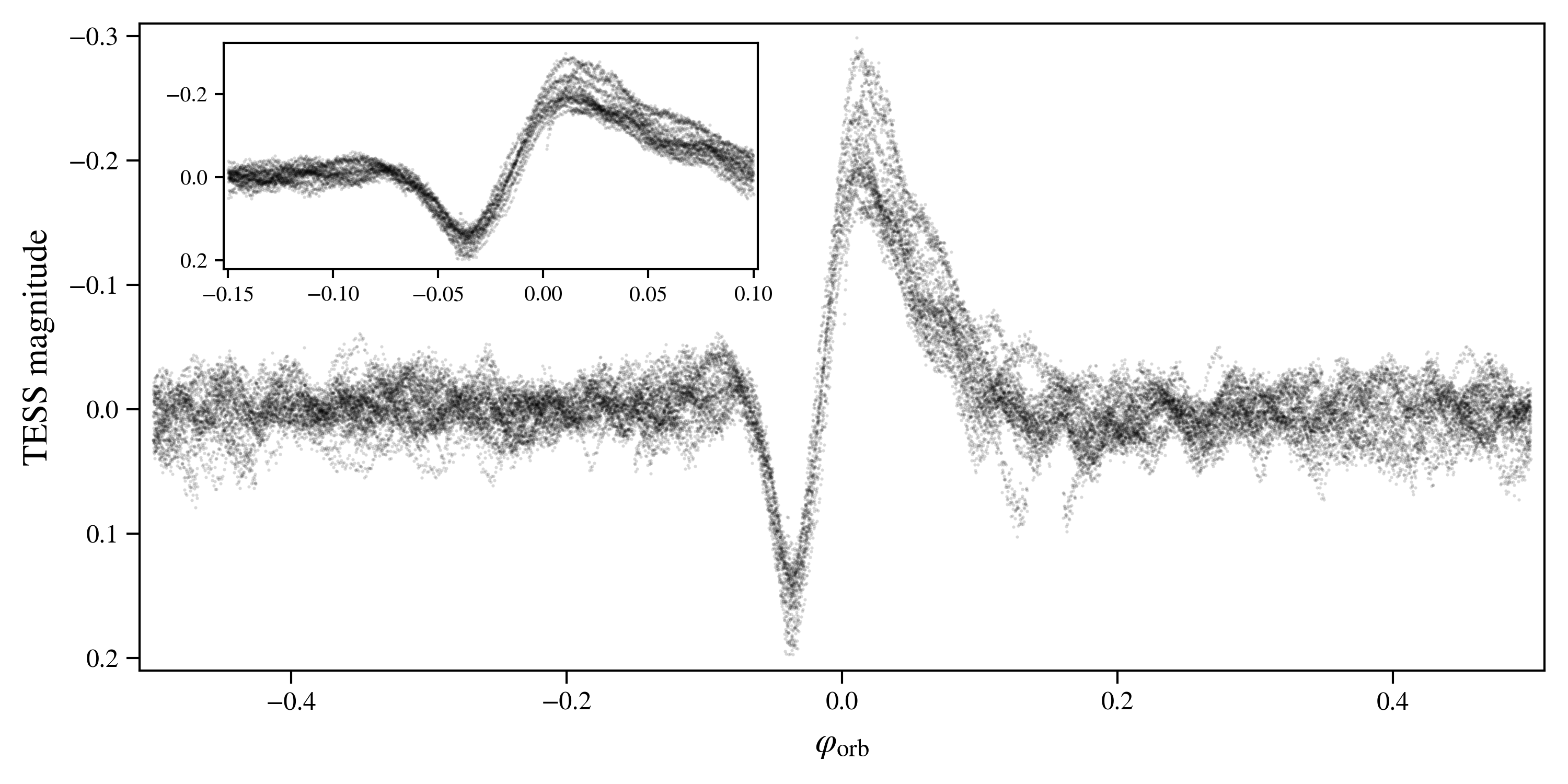}
   \caption{TESS light curve of ExtEV phased with the orbital period of 32.83016\,d. Orbital phase $\varphi_{\rm orb}= 0.0$ corresponds to the time of the periastron passage derived in Sect.~\ref{sect:rv-solution}. The inset shows the light curve near the periastron passage. Stochastic light variations and TEOs can be seen over the entire range of orbital phases.}
   \label{fig:tess_lc}
\end{figure*}

ExtEV is located in the continuous viewing zone of the Transiting Exoplanet Survey Satellite \citep[TESS,][]{2015JATIS...1a4003R} space mission. It was therefore observed during Year 1 (July 2018 -- July 2019), Year 3 (July 2020 -- July 2021) of TESS operation. Presently, the observations of the southern hemisphere scheduled for Year 5 (January -- September 2023) are being completed. In the previous analysis of the TESS data of ExtEV (KS22), we performed photometry of the star on the TESS Full Frame Images obtained during Year 1 and most of the Year 3 observations. In the present paper, we use the same data as they are sufficient for our purposes. The TESS light curve phased with the orbital period is shown in  Fig.\,\ref{fig:tess_lc}. All details of its extraction can be found in KS22.

For the reasons outlined in Sect.~\ref{sect:RLOF}, it is important to check the range of variation and the shape of the light curve of ExtEV in short-wavelength passbands. Unfortunately, we did not find any publicly available light curves in such bands, so we collected near-simultaneous time-series photometry in the Johnson $U$, $B$ and $V$ passbands at three observatories of the Las Cumbres Observatory Global Telescope (LCOGT) network, namely Siding Spring Observatory, Cerro Tololo Inter-American Observatory, and South African Astronomical Observatory. The images of the ExtEV were taken by a total of eight 1-m Ritchey-Chr\'etien telescopes, each equipped with a `Sinistro' CCD Camera\footnote{\url{https://lco.global/observatory/instruments/sinistro/}}. The field of view of the camera is 26.5$\arcmin\,\times\,$26.5$\arcmin$, with an angular resolution of 0.389$\arcsec$ per pixel. The LCOGT data cover an interval of about 68 days from 9 February to 18 April 2022, which was enough to cover just over two orbital cycles of the system. The observations were spread over the whole range of orbital phases with the desired higher density close to periastron passages.

\begin{figure*}
   \centering
   \includegraphics[width=\hsize]{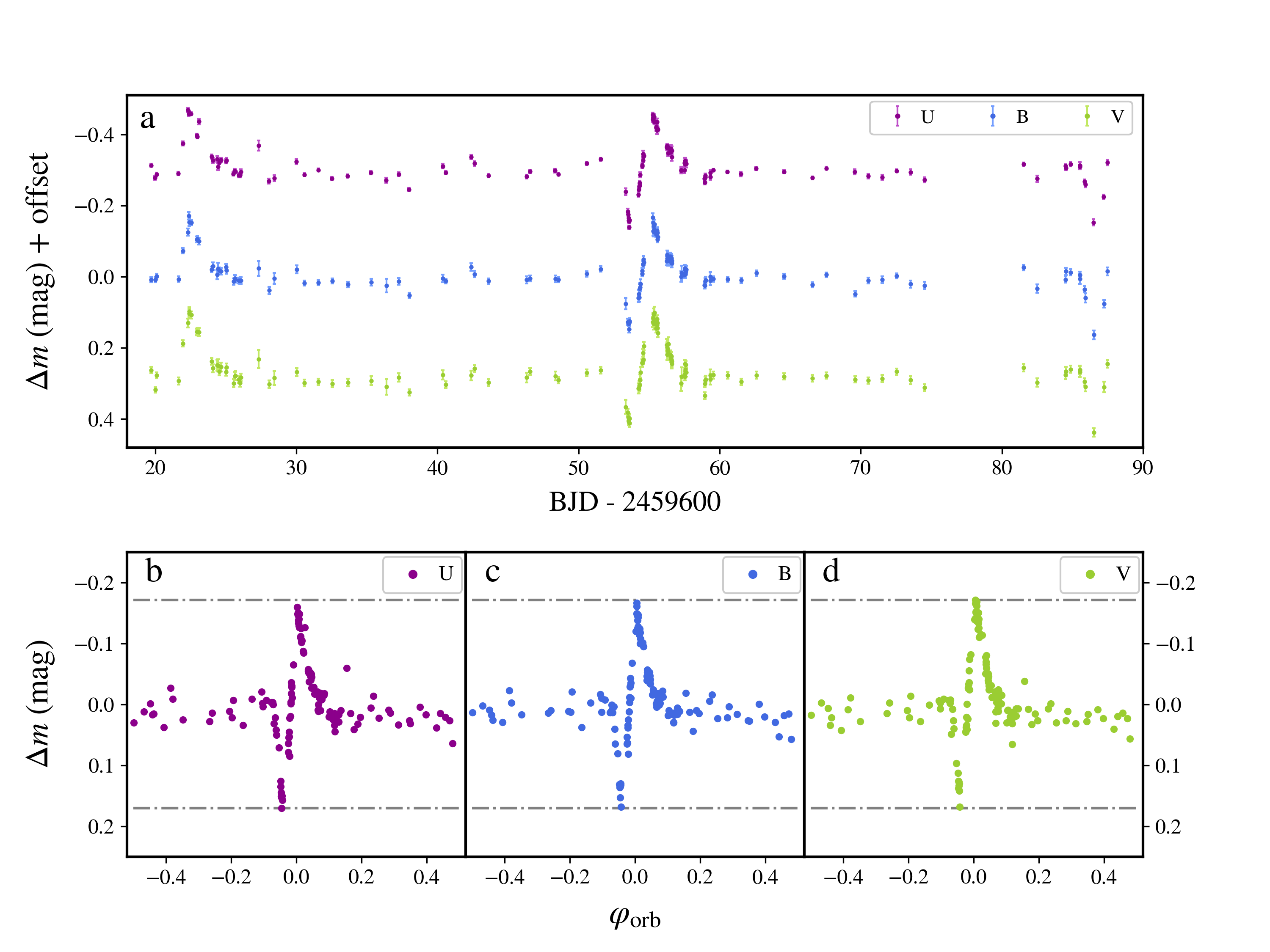}
   \caption{LCOGT light curves of ExtEV in Johnson $U$ (purple dots), $B$ (blue dots), and $V$ (green dots) passbands. Panel (a) shows the light curves. A $\pm0.25$\,mag vertical offset was applied to $U$ and $V$-filter light curves to avoid overlapping. Panels (b), (c), and (d) show the light curves phased with the orbital period. Phase zero corresponds to the time of periastron passage. In addition, we have drawn dash-dotted lines in each of the lower panels to indicate the range of variability during the heartbeat. The plot illustrates that the range is virtually the same in all three passbands.}
   \label{fig:ExtEEV_LCO_2}
\end{figure*}

We performed standard Point Spread Function (PSF) photometry on the collected images using the tools included in \texttt{photutils} Python package\footnote{\url{https://photutils.readthedocs.io/en/stable/}} \citep{larry_bradley_2022_6825092}. By modeling and subtracting the non-uniform background of the LMC, we managed to construct empirical models of the PSF following the procedure described in \cite{2000PASP..112.1360A}, which we then fitted using least-squares to the actual profiles of the ExtEV and several non-variable comparison stars in its vicinity. The resulting light curves are illustrated in Fig.~\ref{fig:ExtEEV_LCO_2}.

\subsection{Time-series optical spectroscopy and radial velocities}\label{sect:Time-series optical spectroscopy and radial velocities}

Using the High-Resolution Spectrograph \citep[HRS,][]{2014SPIE.9147E..6TC} installed on the Southern African Large Telescope \citep[SALT,][]{2022SPIE12186E..0BR}, we collected a series of 18 optical spectra in low-resolution mode with an average resolving power of 14\,000 between 1 October 2021 and 16 March 2022. All SALT/HRS fiber-fed echelle spectra were calibrated and reduced by the HRS MIDAS pipeline\footnote{\url{https://astronomers.salt.ac.za/software/hrs-pipeline/}} 
\citep{2016MNRAS.459.3068K,2017ASPC..510..480K}. The average signal-to-noise ratio (SNR) in the vicinity of the Balmer H$\alpha$ and H$\beta$ lines was around 80 and 50, respectively. Due to the poor quality of two spectra, we ultimately utilized 16 of them in our analysis and normalized them with the automatic tool SUPPNet \citep{2022A&A...659A.199R}.

In the SALT/HRS spectrograph, in addition to the fiber centered on the target, there is a second fiber nearby which is usually used to measure the background. The spectrum measured with this fiber is later used to remove telluric lines from the spectrum of the target. Unfortunately, the ExtEV is located in a crowded field, where an extended and spatially variable emission nebula is present. This makes it difficult to find a location free of any astrophysical sources. We therefore decided to use in our analysis only the spectra from the ExtEV-centered fiber, without correcting for the signal measured by the second fiber, as it contained signal from the emission nebula and nearby fainter stars. We did not fit and remove telluric lines from any of our spectra.

We derived the radial-velocity (RV) curve for ExtEV in the following way. We first selected the spectrum with the highest SNR as the reference spectrum. We then determined the RVs of the remaining spectra relative to the reference spectrum by cross-correlating each of them with the reference spectrum. The cross-correlation function was calculated only in certain wavelength intervals centered on strong absorption lines showing no emission (Table~\ref{table:wavelength-ranges}). In practice, to obtain RV, we fitted a parabola to the cross-correlation function around its global maximum. Then, the position of the peak of the fitted parabola was taken as the measured RV. Knowing these RVs, we Doppler-shifted all spectra so that they could be averaged into a single high-SNR template. We iteratively repeated the procedure described above three times. A newly obtained template served each time as a new reference spectrum. Finally, the RV zero point was calculated by cross-correlating the final reference spectrum with a synthetic spectrum from the BSTAR grid of model stellar atmospheres \citep{2007ApJS..169...83L} with effective temperature $T_{\rm eff}=30\,000$\,K, surface gravity $\log(g/({\rm cm}\,{\rm s}^{-2}))=3.25$\,dex, and half-solar metallicity, convolved with the rotational profile with the projected rotational velocity of 175\,km\,s$^{-1}$ (J21). The resulting RVs, together with the estimated formal errors, are presented in Table~\ref{table:our-rvs}.
\begin{table}
\caption{Wavelength ranges and absorption lines that we used to obtain radial velocities of ExtEV.}
\label{table:wavelength-ranges}      
\centering                         
\begin{tabular}{l l | l l}        
\hline\hline                 
\noalign{\smallskip}
$\lambda$ range (\AA) & Line(s) & $\lambda$ range (\AA) & Line(s)\\
\noalign{\smallskip}
\hline
\noalign{\medskip}
4380\,--\,4407& \ion{He}{i}\,$4388$ &4912\,--\,4948&  \ion{He}{i}\,$4922$\\
4458\,--\,4492& \ion{He}{i}\,$4471$ &5012\,--\,5032&  \ion{He}{i}\,$5016$\\
4632\,--\,4730& \ion{C}{iii}\,$4651$, &5398\,--\,5439& \ion{He}{ii}\,$5411$\\
& \ion{He}{i}\,$4713$,&&\\
& \ion{He}{ii}\,$4686$&&\\
\noalign{\smallskip}
\hline                                   
\end{tabular}
\tablefoot{The given wavelength ranges correspond to the observed (i.e. Doppler-shifted) spectrum, while the wavelengths of the lines correspond to laboratory values.}
\end{table}

\begin{table}
\caption{Barycentric radial velocities $v_{\rm rad}$ of the primary component of ExtEV extracted from the SALT/HRS spectra.}
\label{table:our-rvs}      
\centering                         
\begin{tabular}{c c | c c}        
\hline\hline                 
\noalign{\smallskip}
BJD & $v_{\rm rad}$ & BJD & $v_{\rm rad}$\\
$-$ 2\,450\,000 &(km\,s$^{-1}$) & $-$ 2\,450\,000 & (km\,s$^{-1}$)\\
\noalign{\smallskip}
\hline
\noalign{\medskip}
 9459.6291   &   $+$363.2(11) & 9525.4729   &   $+$359.0(11)\\
 9461.6065   &   $+$355.9(27) & 9528.4326   &   $+$337.6(14)\\
 9493.5252   &   $+$375.9(40) & 9530.4650   &   $+$294.3(26)\\
 9494.5583   &   $+$348.3(28) & 9550.3652   &   $+$208.3(34)\\
 9512.4829   &   $+$238.8(25) & 9559.3702   &   $+$357.5(16)\\
 9518.4816   &   $+$211.8(12) & 9560.3423   &   $+$337.4(23)\\
 9521.4845   &   $+$247.9(32) & 9628.3107   &   $+$300.5(40)\\
 9523.4583   &   $+$325.9(19) & 9630.2836   &   $+$280.0(30)\\
\noalign{\smallskip}
\hline                                   
\end{tabular}
\end{table}

\section{SED fitting}\label{sect:sed-fitting}

\begin{table*}
\caption{Multi-color photometry of ExtEV used by us to perform SED fitting described in Sect.~\ref{sect:sed-fitting}.}
\label{table:mags-for-seds}      
\centering                         
\begin{tabular}{l c r l l}        
\hline\hline                 
\noalign{\smallskip}
Mission\,/\,Survey & Passband & $\lambda_{\rm eff}$\,($\mu{\rm m}$)$^\ast$ & Magnitude & Reference\\
\noalign{\smallskip}
\hline
\noalign{\medskip}
Swift/UVOT & UVW2 & 0.208 & 14.22(3) & J21\\
 & UVM2 & 0.225 & 14.24(4) & J21\\
 & UVW1 & 0.268 & 13.92(4) & J21\\
\noalign{\medskip}
GALEX & NUV & 0.230 & 15.62(5) & This work\\
\noalign{\medskip}
XMM/OM & UVW1 & 0.268 & 13.3248(14) & \cite{2021yCat.2370....0P}\\
\noalign{\medskip}
{The Magellanic Clouds} & $U$ & 0.355 & 12.63(4) & \cite{2004AJ....128.1606Z}\\
{Photometric Survey} & $B$ & 0.437 & 13.62(11) & \cite{2004AJ....128.1606Z}\\
& $I$ & 0.857 & 13.28(8) & \cite{2004AJ....128.1606Z}\\
\noalign{\medskip}
$UBVR$ CCD Survey of & $U$ & 0.355 & 12.830(17) & \cite{2002yCat.2236....0M}\\
the Magellanic Clouds & $B$ & 0.437 & 13.670(14) & \cite{2002yCat.2236....0M}\\
& $V$ & 0.547 & 13.560(10) & \cite{2002yCat.2236....0M}\\
& $R$ & 0.670 & 13.430(22) & \cite{2002yCat.2236....0M}\\
\noalign{\medskip}
Gaia DR3 & $G_{\rm BP}$ & 0.504 & 13.504(6) & \cite{2022yCat.1355....0G}\\
& $G$ & 0.582 & 13.423(3) & \cite{2022yCat.1355....0G}\\
& $G_{\rm RP}$ & 0.762 & 13.227(7) & \cite{2022yCat.1355....0G}\\
\noalign{\medskip}
APASS DR10 & $B$ & 0.437 & 13.678(9) & \cite{2019JAVSO..47..130H}\\
& $g'$ & 0.472 & 13.542(7) & \cite{2019JAVSO..47..130H}\\
& $V$ & 0.547 & 13.530(13) & \cite{2019JAVSO..47..130H}\\
& $r'$ & 0.620 & 13.568(21) & \cite{2019JAVSO..47..130H}\\
& $i'$ & 0.767 & 13.604(22) & \cite{2019JAVSO..47..130H}\\
\noalign{\medskip}
OGLE & $V$ & 0.547 & 13.7720(10) & \cite{2022yCat..22590016W}\\
& $I_{\rm C}$ & 0.783 & 13.4730(10) & \cite{2022yCat..22590016W}\\
\noalign{\medskip}
2MASS & $J$ & 1.235 & 12.992(23) & \cite{2012yCat.2281....0C}\\
& $H$ & 1.662 & 12.847(25) & \cite{2012yCat.2281....0C}\\
& $K_{\rm s}$ & 2.159 & 12.699(24) & \cite{2012yCat.2281....0C}\\
\noalign{\medskip}
Spitzer/IRAC & Ch1 & 3.508 & 12.41(3) & \cite{2009AJ....138.1003B}\\
& Ch2 & 4.437 & 12.30(3) & \cite{2009AJ....138.1003B}\\
& Ch3 & 5.628 & 12.14(7) & \cite{2009AJ....138.1003B}\\
& Ch4 & 7.589 & 11.87(9)& \cite{2009AJ....138.1003B}\\
\noalign{\medskip}
AllWISE & W1 & 3.353 & 12.416(24) & \cite{2014yCat.2328....0C}\\
& W2 & 4.603 & 12.377(22) & \cite{2014yCat.2328....0C}\\
& W3 & 11.561 & 9.90(7) & \cite{2014yCat.2328....0C}\\
& W4 & 22.088 & 6.20(9) & \cite{2014yCat.2328....0C}\\
\noalign{\smallskip}
\hline                                   
\end{tabular}
\tablefoot{$^\ast$ Effective wavelengths of passbands as calculated by VOSA VO SED Analyzer \citep{2008A&A...492..277B}.}
\end{table*}

Although J21 performed an SED fit to the selected photometry available for ExtEV, we decided to repeat this task because of two reasons. Firstly, as we argue below, ExtEV shows an infrared (IR) excess, and J21 did not exclude the near-IR (NIR) bands from their fit, which may have biased the result. Secondly, due to the high effective temperature of the primary component, every single measurement in the ultraviolet (UV) is particularly important. For ExtEV, we found several sources of such measurements in the literature (Table~\ref{table:mags-for-seds}). We also utilize the fact that the system has been observed by the Galaxy Evolution Explorer \citep[GALEX,][]{2007ApJS..173..682M} satellite in the near-UV (NUV) band, although GALEX photometry for ExtEV has not been published. We therefore extracted our own photometry from the `tile' in the NUV passband from the GALEX\,GR6/7 data release \citep{2014yCat.2335....0B} as follows. We downloaded a tile named `LMC\_39774\_0422' from the MAST portal\footnote{\url{https://galex.stsci.edu/GR6/}}, on which we performed circular aperture photometry of ExtEV using the \texttt{photutils} tool. We estimated the mean background signal at the ExtEV position using a nearby region without any point sources. We then converted the measured signal to magnitude using a tool provided at GALEX website\footnote{\url{https://asd.gsfc.nasa.gov/archive/galex/FAQ/counts_background.html}}.
A full list of ExtEV magnitudes with references can be found in Table~\ref{table:mags-for-seds}.
\begin{figure}
   \centering
   \includegraphics[width=\hsize]{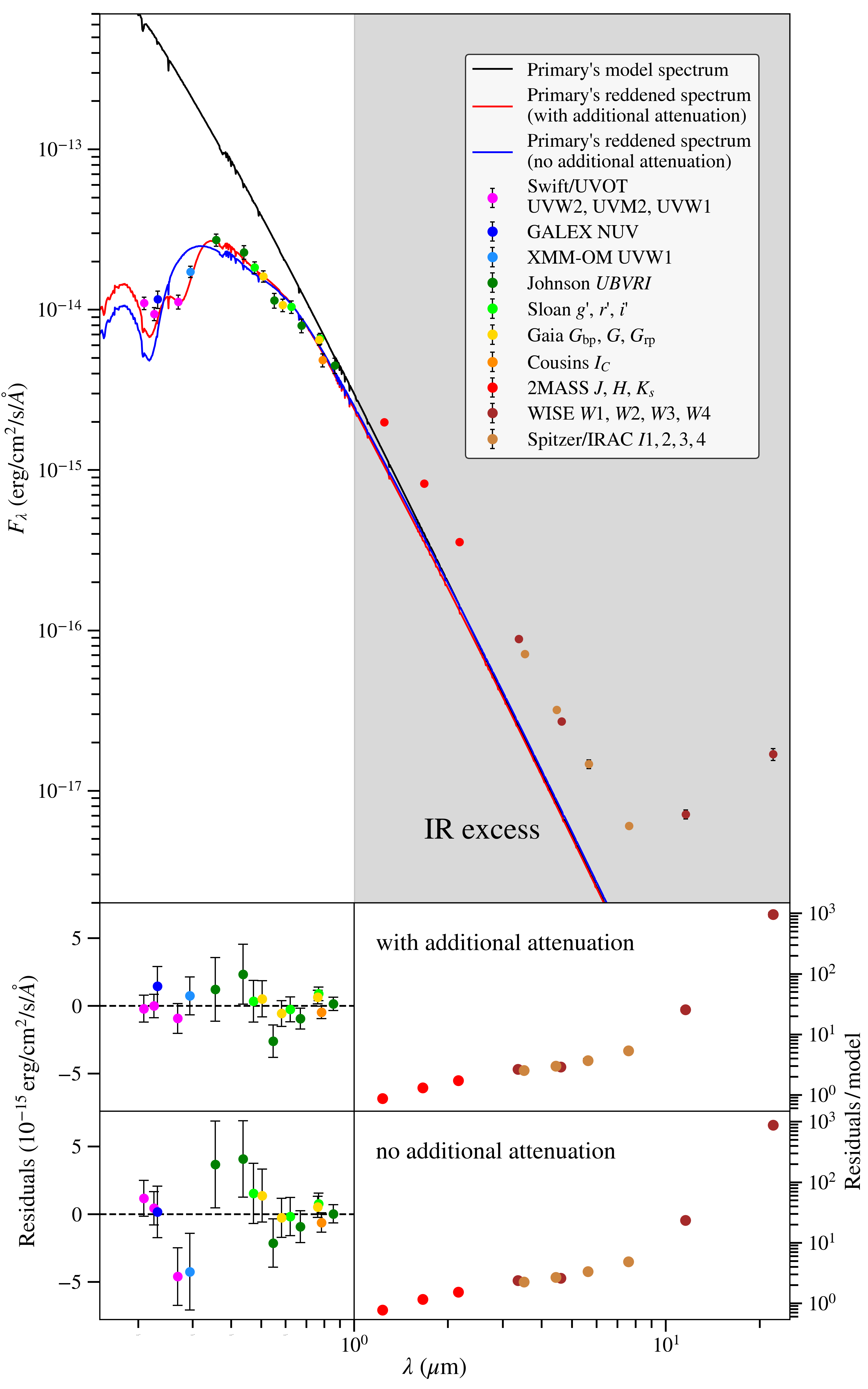}
   \caption{Results of SED fitting to the multi-band photometry of ExtEV. The upper panel shows the absolute fluxes calculated from magnitudes given in Table~\ref{table:mags-for-seds} (dots are color-coded according to the legend). The model spectrum of the primary component is shown with a black line, its reddened versions with a red line (with additional attenuation) and a blue line (without it). The infrared excess is clearly visible and indicated by the shaded area. The middle and lower panels show the residuals from the best fit, with and without the additional attenuation, respectively. For residuals presented on the left (UV and optical range), the ordinate axes are linearly scaled while the residuals in the IR range are normalized by the model values and logarithmically scaled.}
   \label{fig:double-dust-sed-fit_v3}
\end{figure}

To estimate the radius of the primary component, $R_1$, and consequently its luminosity, $L_1$, we performed the following model fitting to the observed SED. First, we converted all magnitudes presented in Table~\ref{table:mags-for-seds} to absolute spectral flux densities using the VOSA\,7.5 online tool\footnote{\url{http://svo2.cab.inta-csic.es/theory/vosa/}} \citep{2008A&A...492..277B}. If a given photometric passband had more than one measurement in Table~\ref{table:mags-for-seds}, we averaged the corresponding fluxes. Due to the IR excess of ExtEV, we considered only passbands with effective wavelengths shorter than the $J$ filter. We then assumed that the distance to the ExtEV is equal to the distance to the LMC \citep[49.59\,kpc,][]{2019Natur.567..200P} and that the absolute spectrum of the primary component corresponds to the ATLAS9 model atmosphere \citep{2003IAUS..210P.A20C} with the parameters $T_{\rm eff}=30\,000\,$K, $\log(g/({\rm cm}\,{\rm s}^{-2}))=3.5$\,dex, ${\rm [Fe/H]}=-0.5$\,dex, and solar-scaled chemical composition (Fig.~\ref{fig:double-dust-sed-fit_v3}, black line). The first two parameters were obtained by J21, while the last one is close to the mean metallicity of the LMC \citep[which is about $-0.42$\,dex,][]{2021MNRAS.507.4752C}. Based on the adopted synthetic spectrum and the extinction law of the interstellar medium (ISM) of \cite{1999PASP..111...63F}, implemented in the \texttt{extinction} Python package \citep{barbary_kyle_2016_804967}, we were able to calculate the theoretical SED with the \texttt{sedpy} tool \citep{johnson_benjamin_d_2021_4582723}. We performed a Bayesian fit of the theoretical values to the observed ones, using the standard maximum-likelihood estimation method. However, we took the so-called noise nuisance parameter, $\xi$, into account when calculating the total flux density errors, so they were given by $\sigma_{\rm total}^2=\sigma_{\rm phot}^2+\mathcal{M}^2\exp (2\ln \xi)$, where $\sigma_{\rm phot}$ refers to the reported photometric error and $\mathcal{M}$ stands for the model SED value. The purpose of this procedure is twofold. Firstly, photometric errors are usually underestimated. Secondly, in addition to the `heartbeat' feature, ExtEV also exhibits TEOs and stochastic variations in its light curve, resulting in some intrinsic scatter in the photometric measurements\footnote{Fortunately, the probability of any of the above measurements coinciding with an orbital phase close to the periastron passage is relatively small. Indeed, in Fig.~\ref{fig:double-dust-sed-fit_v3} we do not observe any strong outliers, except those affected by the IR excess.} collected in Table~\ref{table:mags-for-seds}. When fitting the theoretical SED, the following parameters were considered free: the total extinction in the $V$ band, $A_V$, the ratio of total to selective extinction, $R_V$, $R_1$, and $\ln\xi$. Once the optimal solution was identified, we estimated the errors of the fitted parameters using the affine invariant Markov Chain Monte Carlo (MCMC) method \citep{2010CAMCS...5...65G}, using the implementation provided by the \texttt{emcee} package \citep{2013PASP..125..306F}.

\begin{figure}
   \centering
   \includegraphics[width=\hsize]{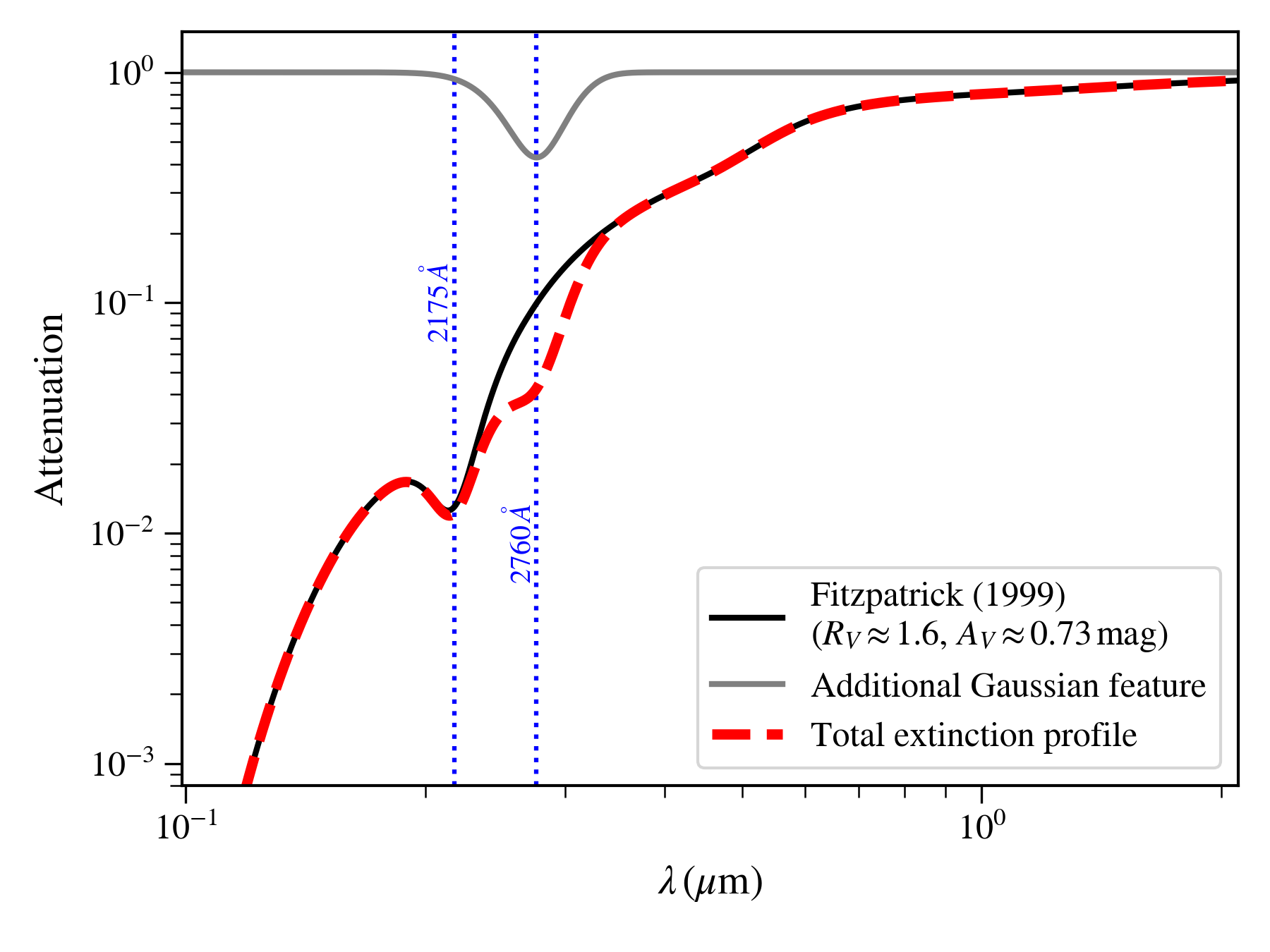}
   \caption{Extinction law resulting from the best fit to observed SED of ExtEV. The black curve refers to the interstellar dust extinction function of \cite{1999PASP..111...63F} with the parameters given in the legend. The additional extinction feature is shown with the gray line, while the red dashed curve corresponds to the total extinction function (the product of black and gray curves). With a pair of blue vertical dotted lines, we have marked the positions of the two local minima in the total extinction function. More details can be found in the main text.}
   \label{fig:double-extinction_v3}
\end{figure}

Interestingly, fitting the SED using the extinction law of  \cite{1999PASP..111...63F} results in `U'-shaped residuals in the UV region and the $U$ and $B$ passbands (Fig.~\ref{fig:double-dust-sed-fit_v3}, bottom panel) regardless of the values of $R_V$ and $A_V$. We also find that this situation cannot be avoided by choosing different prescriptions for the ISM extinction function. This behavior of the residuals suggests the presence of an additional source of attenuation that is not accounted for by the ISM extinction laws. Therefore, to consider the possible existence of an additional extinction component, we multiplied the interstellar extinction law by the following Gaussian term, $1-\alpha_{\rm G}\exp[(\lambda-\lambda_{\rm G})^2/2\sigma_{\rm G}^2]$, where $\alpha_{\rm G}$, $\lambda_{\rm G}$, and $\sigma_{\rm G}$ stand for the amplitude, mean, and standard deviation of the Gaussian profile, respectively (Fig.~\ref{fig:double-extinction_v3}). The effects of this fitting procedure are shown in Fig.~\ref{fig:double-dust-sed-fit_v3}, while the optimized parameters can be found in Table~\ref{table:sed-fit}.

\begin{table}
\caption{Optimized parameters of ExtEV, resulting from fitting SED to the data from Table~\ref{table:mags-for-seds}.}
\label{table:sed-fit}      
\centering                         
\begin{tabular}{l r}        
\hline\hline                 
\noalign{\smallskip}
Parameter & Optimized value with error\\
\noalign{\smallskip}
\hline  
\noalign{\medskip}
$A_V$\,(mag) & $0.73^{+0.18}_{-0.15}$\\
\noalign{\smallskip}
$R_V$ & $1.60^{+0.29}_{-0.21}$\\
\noalign{\smallskip}
$R_1$\,(R$_\sun$) & $30.0^{+2.4}_{-1.9}$\\
\noalign{\smallskip}
$\lambda_{\rm G}$\,($\AA$) & $2760^{+150}_{-290}$\\
\noalign{\smallskip}
$\alpha_{\rm G}$ & $0.65^{+0.18}_{-0.17}$\\
\noalign{\smallskip}
$\sigma_{\rm G}$\,($\AA$) & $210^{+300}_{-180}$\\
\noalign{\smallskip}
$\ln\xi$ & $-2.48^{+0.25}_{-0.22}$\\
\noalign{\smallskip}
\hline
\noalign{\smallskip}
$\log(L_1$/L$_\sun)^\ast$ & $5.82^{+0.07}_{-0.05}$\\
\noalign{\smallskip}
\hline                                   
\end{tabular}
\tablefoot{$^\ast$ Calculated from the obtained value of $R_1$ assuming $T_{\rm eff}=30\,000$\,K for the primary component.}
\end{table}

\begin{figure*}
   \sidecaption
   \includegraphics[width=12cm]{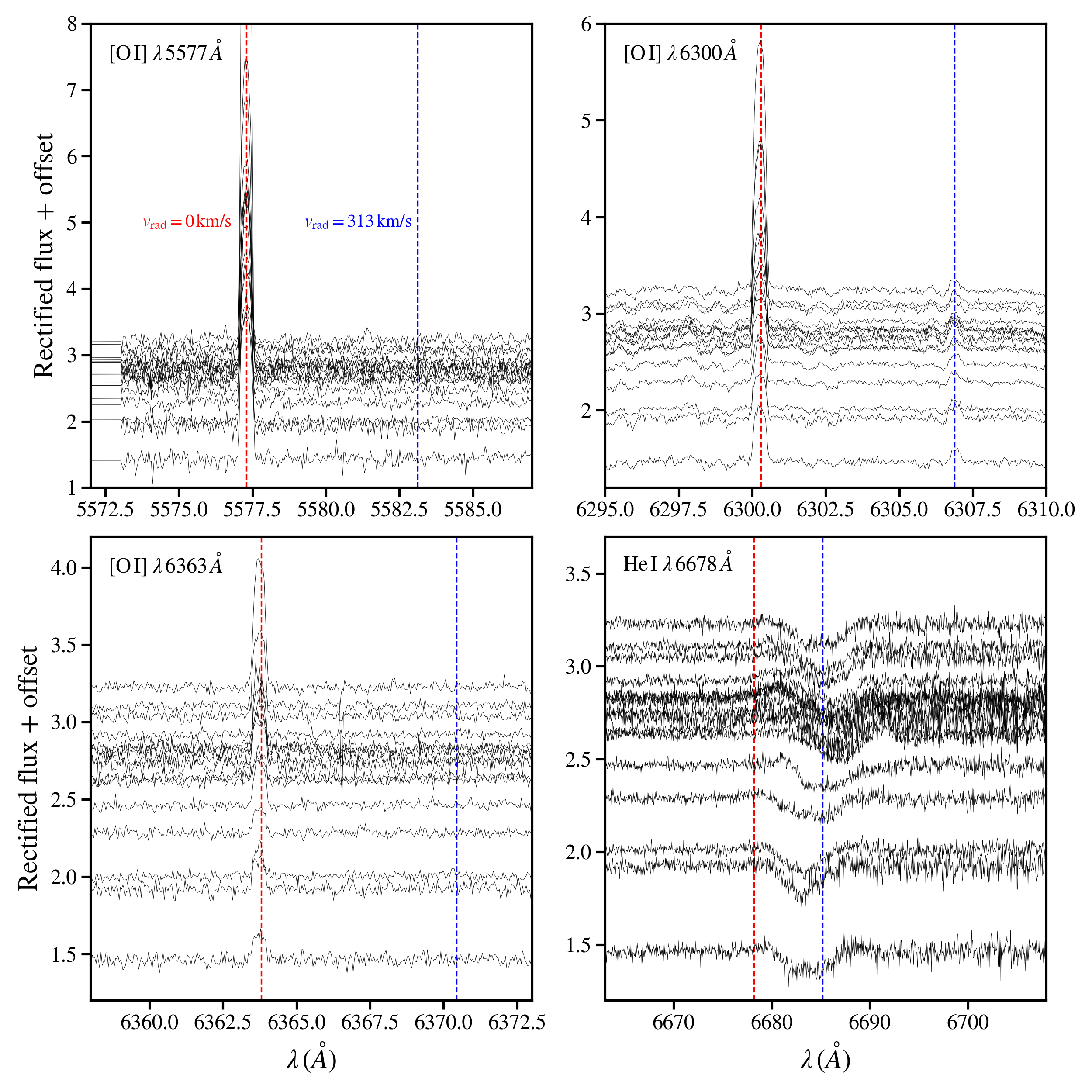}
   \caption{Small sections of the SALT spectra showing [\ion{O}{i}] emission lines at their laboratory wavelengths but not in the rest frame of ExtEV. The rest-frame wavelengths of the lines shown are indicated in each panel in the top left corner. The spectra are shifted vertically according to the orbital phase. The vertical red and blue dashed lines correspond to RVs of 0 (geocoronal emission) and 313\,km\,s$^{-1}$ (systemic velocity of ExtEV), respectively. For comparison, the bottom right panel shows the \ion{He}{i} absorption line originating from the photosphere of the primary component of ExtEV.}
   \label{fig:OI_lines}
\end{figure*}

The best-fit value of $R_1$ is 30\,R$_\sun$, which is significantly higher than the value obtained by J21 (their $R_1\approx 24$\,R$_\sun$). This also implies that the value of $L_1$ is higher than that obtained by J21, amounting to about $\log(L_1$/L$_\sun)=5.82$. The location of the ExtEV's primary component in the HR diagram will be discussed in Sect.~\ref{sect:mass-estimation}. The rather unusual value of $R_V=1.6$ and the additional attenuation component in the total extinction function can be attributed to the presence of dust in the vicinity of the ExtEV with grain sizes and chemical composition significantly different from those characterizing the Galactic ISM \citep[e.g.][]{2016P&SS..133...36N}. Since polycyclic aromatic hydrocarbon (PAH) molecules are responsible for the well-known absorption maximum around $2175\,\AA$ \citep[e.g.][]{2017ApJ...836..173B}, depending on the conditions of the medium and their mixture, they can give rise to additional relatively broad absorption bands in the UV range \citep{2004A&A...426..105M, 2011ApJ...742....2S}. Perhaps the additional attenuation factor seen in the ExtEV SED comes from PAHs or other molecules and mineral grains not typical of the ISM.


\begin{figure}
   \centering
   \includegraphics[width=\hsize]{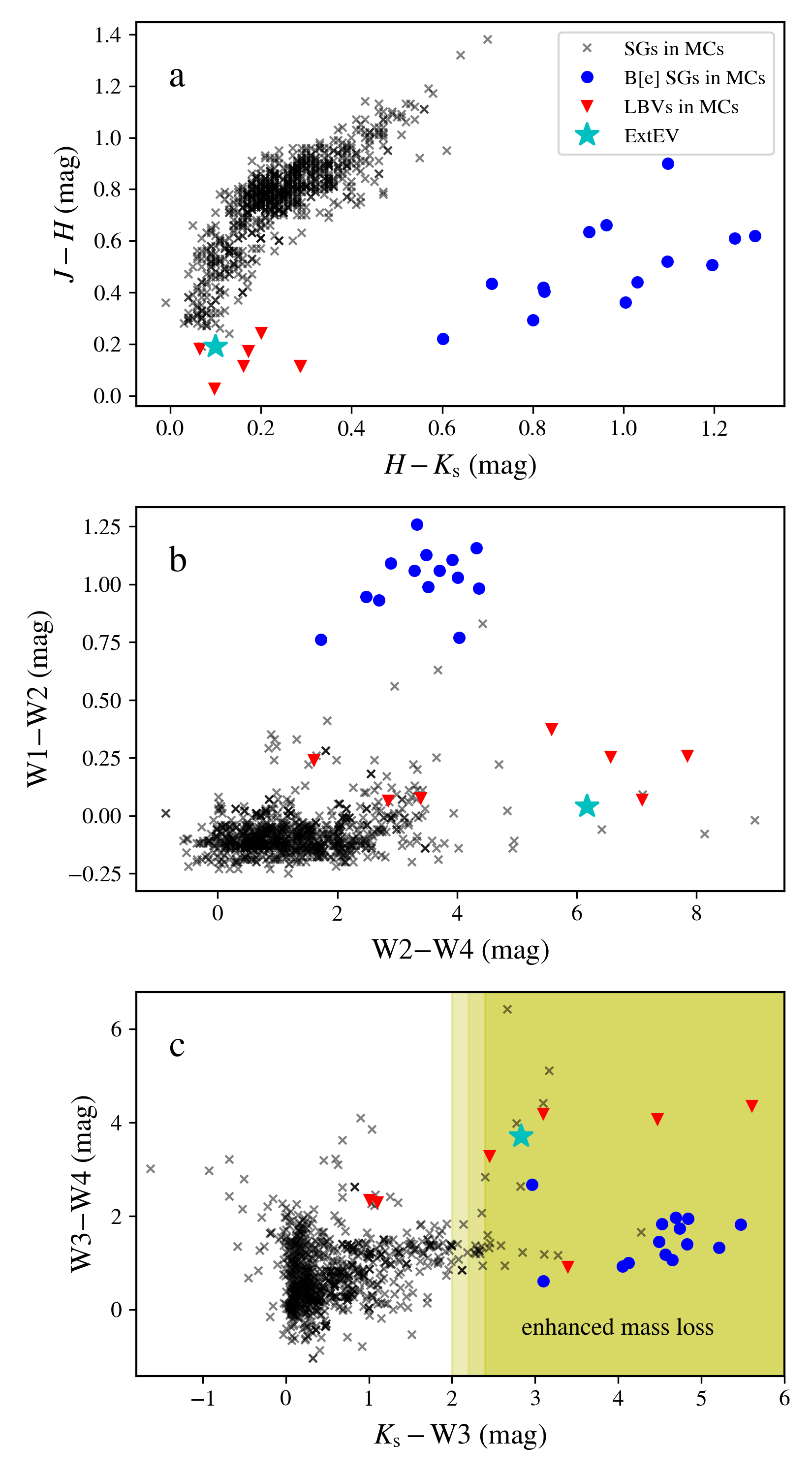}
   \caption{Infrared color-color diagrams for ExtEV. The position of ExtEV is marked by the cyan star in each panel. The blue dots and red triangles correspond to the B[e] SGs and LBVs located in the Magellanic Clouds, respectively. Their photometric properties were adopted from \cite{2019Galax...7...83K}. Black crosses refer to the sample of late-type SGs selected by \cite{2015A&A...578A...3G}. The shaded region in panel (c) shows the area with color index $({\rm K}_{\rm s}-{\rm W}3)\gtrsim2$\,mag, where SGs with enhanced mass loss are suspected to be located.}
   \label{fig:ir-photometry}
\end{figure}

\section{Properties of the ExtEV system and its primary component}\label{sect:results}

\subsection{Non-B[e] status of ExtEV}\label{sect:Non-B[e] status of ExtEV}

B[e] supergiants are characterized by the presence of two components in their intense stellar wind: a `fast wind' located around the poles and a `slow wind' in the form of an outflowing disk emanating at a certain angle from the equatorial region \citep[e.g.][]{2007A&A...463..627K}. In addition to the strong emission in the hydrogen lines, the primary distinguishing feature of B[e] SGs is the presence of forbidden [\ion{O}{i}] lines at 5577, 6300, and 6363\,$\AA$ \citep[e.g.][]{1998A&A...340..117L}, which form in the immediate vicinity of the star. Other distinguishing features include emission in the \ion{Fe}{ii} and [\ion{Fe}{ii}] lines, as well as strong molecular CO emission and a significant NIR excess, which originates from relatively hot circumstellar dust with temperatures of the order of 1000\,K \citep{2019Galax...7...83K}.

Although J21 reported the presence of some of the aforementioned emission lines in their spectra and claimed that ExtEV is a B[e] SG with a disk, this emission disappeared in some of their spectra (their Fig.~16). In none of our spectra from SALT/HRS, did we observe the presence of the emission lines [\ion{O}{i}], \ion{Fe}{ii}, and [\ion{Fe}{ii}]. This may suggest that the detections made by J21 were unlikely to involve ExtEV. Some of these could come, for example, from cosmic rays or inadequate removal of nebular and geocoronal emission lines. The absence of the three strongest [\ion{O}{i}] lines in our spectra is presented in Fig.~\ref{fig:OI_lines}. For reasons explained in Sect.~\ref{sect:Time-series optical spectroscopy and radial velocities}, only [\ion{O}{i}] lines resulting from geocoronal emission are present at $v_{\rm rad}=0$\,km\,s$^{-1}$. In the vicinity of $v_{\rm rad}=313\,$km\,s$^{-1}$, the systemic velocity of the ExtEV (Sect.~\ref{sect:rv-solution}), no significant [\ion{O}{i}] emission is observed over a period of about six months covering different orbital phases. Hence, we would like to conclude that the SALT/HRS spectra of ExtEV do not show the typical spectral features of B[e] SGs.

Analysis of the photometric data presented in Sect.~\ref{sect:sed-fitting} indicates that ExtEV shows a strong IR excess, particularly in the mid-IR (MIR) range (W3 and W4 WISE passbands). Therefore, we decided to investigate the infrared properties of ExtEV by reproducing the NIR and WISE color-color diagrams from \citet[][their Fig.~6]{2019Galax...7...83K}. The result of this analysis is shown in Fig.~\ref{fig:ir-photometry}. It becomes apparent that ExtEV has markedly different IR properties compared to B[e] SGs and instead falls within the range typical of luminous blue variables \citep[LBVs, e.g.][]{2017RSPTA.37560268S}. This implies that there is no hot dust in the vicinity of ExtEV, which usually accumulates relatively close to the central SG. On the contrary, the clear excess in the MIR (Fig.~\ref{fig:ir-photometry}b) suggests the presence of cooler dust, probably surrounding the entire binary system at a considerable distance from both components. In the case of SGs, the (${\rm W}3-{\rm W}4$) -- ($K_{\rm s}-{\rm W}3$) color-color diagram may serve as an indirect indicator of the circumstellar dust temperature and mass loss rate, respectively \citep[][their Fig.~15]{2015A&A...578A...3G}. In the case of ExtEV, $(K_{\rm s}-{\rm W}3)\approx 2.8$~mag, indicating that the system is undergoing enhanced mass loss (Fig.~\ref{fig:ir-photometry}c), a feature common to both LBVs and B[e] SGs.

Given all of the above, we can rule out the possibility that ExtEV is a B[e] SG. This conclusion is further strengthened by the analysis presented in Sect.~\ref{sect:disk}. Whatever the conditions in the vicinity of the primary component of ExtEV are, with the exception of the increased rate of mass loss, they differ from those typical of B[e] SGs.

\subsection{Estimating the mass of the primary component}\label{sect:mass-estimation}

\begin{table*}
\caption{Initial parameters with their distributions which were used to synthesize 1000 evolutionary tracks of the primary component using MESA software.}
\label{table:mesa-distributions}      
\centering                         
\begin{tabularx}{\textwidth}{X c X}
\hline\hline                 
\noalign{\smallskip}
Parameter & Distribution$^\ast$ & Comment\\
\noalign{\smallskip}
\hline    
\noalign{\medskip}
Initial mass at ZAMS & $M_{\rm 1,ZAMS}$/M$_\sun\sim\mathcal{L}(25,60)$ & Log-uniform distribution was used to maintain a nearly uniform density of evolutionary tracks in the HR diagram.\\
\noalign{\smallskip}
Solar-scaled metallicity & ${\rm [Fe/H]}\sim\mathcal{N}(-0.42,0.04)$ & Parameters of the normal distribution were taken from \cite{2021MNRAS.507.4752C} to be consistent with the stellar metallicity distribution in the LMC.\\
\noalign{\smallskip}
Initial angular velocity of rotation normalized by its critical velocity & $\Omega_{\rm rot}/\Omega_{\rm crit}\sim\mathcal{U}(0.1,0.7)$ & To account for the significant rotation rates of young massive stars, which trigger several mixing mechanisms in stellar interiors.\\
\noalign{\smallskip}
Parameter of exponential overshooting from the convective core & $f_{\rm ov}\sim\mathcal{U}(0.01,0.05)$ & The range adopted after \cite{2017ApJ...835..290O}.\\
\noalign{\smallskip}
Scaling factor of wind mass-loss rate & $S_{\rm wind}\sim\mathcal{U}(0.5,5)$ & The actual mass-loss rate was obtained by multiplying its nominal value by $S_{\rm wind}$.\\
\noalign{\smallskip}
\hline                                   
\end{tabularx}
\tablefoot{$^\ast$ $\mathcal{U}(a,b)$ -- uniform distribution on $(a,b)$ range, $\mathcal{L}(a,b)$ -- log-uniform distribution on $(a,b)$ range, $\mathcal{N}(a,b)$ -- normal distribution with $a$ and $b$ corresponding to its mean and standard deviation, respectively.}
\end{table*}

\begin{figure*}
   \centering
   \includegraphics[width=\hsize]{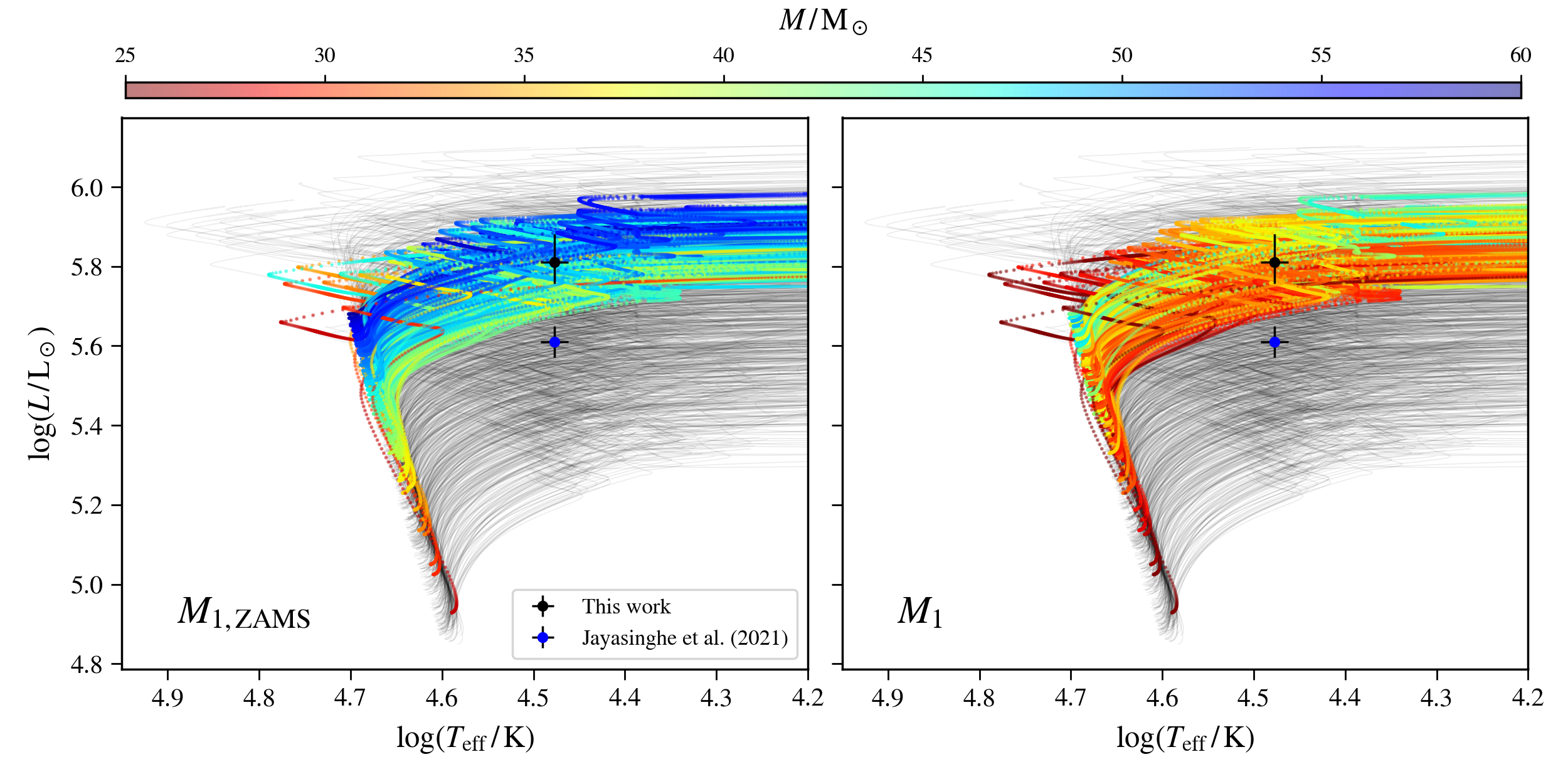}
   \caption{HR diagrams summarizing our simulations described in Sect.~\ref{sect:mass-estimation}. Black and blue dots show the location of the primary component of ExtEV, according to this paper and the study of J21, respectively. The gray curves show evolutionary tracks that never approach the position of the primary component determined in our study. Conversely, multicolored tracks correspond to those that are within the error box of the ExtEV (black dot) at some phase of the evolution. The color bar above the HR diagrams reflects $M_{\rm 1,ZAMS}$ (left panel) and $M_1$ (right panel) and is common to both panels.}
   \label{fig:HRD}
\end{figure*}

Having estimated the effective temperature of the primary component of ExtEV (J21) and its luminosity (Sect.~\ref{sect:sed-fitting}), we were able to locate it in the HR diagram (Fig.~\ref{fig:HRD}). In this way, we could estimate its initial mass at zero-age MS (ZAMS), $M_{\rm 1,ZAMS}$, and its current mass, $M_1$, by examining the evolutionary tracks consistent with the location of the primary component in the HR diagram. To do so, we used the Modules for Experiments in Stellar Astrophysics \citep[MESA; version r22.11.1,][]{Paxton2011, Paxton2013, Paxton2015, Paxton2018, Paxton2019, Jermyn2023} stellar structure and evolution code, compiled with the MESA Software Development Kit \citep[version 21.4.1,][]{2021zndo...5802444T}. Since we were not interested in the precise modeling of the evolution of ExtEV as a binary system, but only in the range of probable $M_{\rm 1,ZAMS}$ and $M_1$, we synthesized a series of 1000 single-star evolutionary tracks that could potentially intersect the current position of primary component in the HR diagram within 1-$\sigma$ error bars. The evolution of massive stars is very sensitive not only to $M_{\rm ZAMS}$ and its metallicity, but also to the rotation velocity divided by its critical value $\Omega_{\rm rot}/\Omega_{\rm crit}$ in the ZAMS and the mass-loss rate due to the line-driven stellar wind, $\dot{M}_{\rm wind}$. In order to realistically capture the richness of possible evolutionary tracks, we computed them with their initial parameters drawn from the distributions presented in Table~\ref{table:mesa-distributions}. While the ranges of the first four parameters presented in this table require no further comment, the last parameter necessitates a deeper justification. For massive stars, the most commonly used formula for $\dot{M}_{\rm wind}$ is the one given by \cite{2001A&A...369..574V}. Our calculations in MESA also used this formula, but we noted that it did not account for any effects related to binarity. On the one hand, there is a growing evidence \citep[e.g.][]{2023A&A...676A.109B} that the aforementioned analytical prescription for $\dot{M}_{\rm wind}$ may in some cases overestimate the mass-loss rate by a factor of 2\,--\,3. Hence, we set the lower limit of $S_{\rm wind}$ to 0.5 to account for the situation where the primary component evolved effectively as a single star. On the other hand, this study provides strong evidence indicating an enhanced mass-loss rate that is ongoing in ExtEV. Given the current state of the primary component, it is not excluded that in the past $\dot{M}_{\rm wind}$ was also larger than typical (e.g.~due to the presence of a close massive companion). We have therefore allowed for situations in which the actual $\dot{M}_{\rm wind}$ may be up to five times more intense than predicted by the \cite{2001A&A...369..574V} formula. The remaining MESA parameters and the set of `physics switches' were identical to those described by \citet[][their Sect.~3.2.2]{2023A&A...671A..22K}.

Looking at the left panel of Fig.~\ref{fig:HRD} one can see that the initial mass of the primary component is most likely in the range between 27 and 55\,M$_\sun$. A more precise value cannot be given because it would be necessary to know the past evolution of ExtEV. The current value of $M_1$ is a separate issue, given the strong stellar wind in this mass range. The right panel of Fig.\,\ref{fig:HRD} suggests that the primary component has currently a mass between 25 and 45\,M$_\sun$. For simplicity, we assume $M_1=35$\,M$_\sun$ in the subsequent sections of this paper and discuss the potential impact of changing this value on our findings. It is also difficult to make a definitive conclusion about the evolutionary status of the ExtEV, as our experiment indicates that the primary component may either be still at the MS, close to the terminal-age MS (TAMS), or has already ignited helium in its core. So it may be a post-MS star that crosses the Hertzsprung gap.

\begin{figure*}
   \sidecaption
   \includegraphics[width=12cm]
{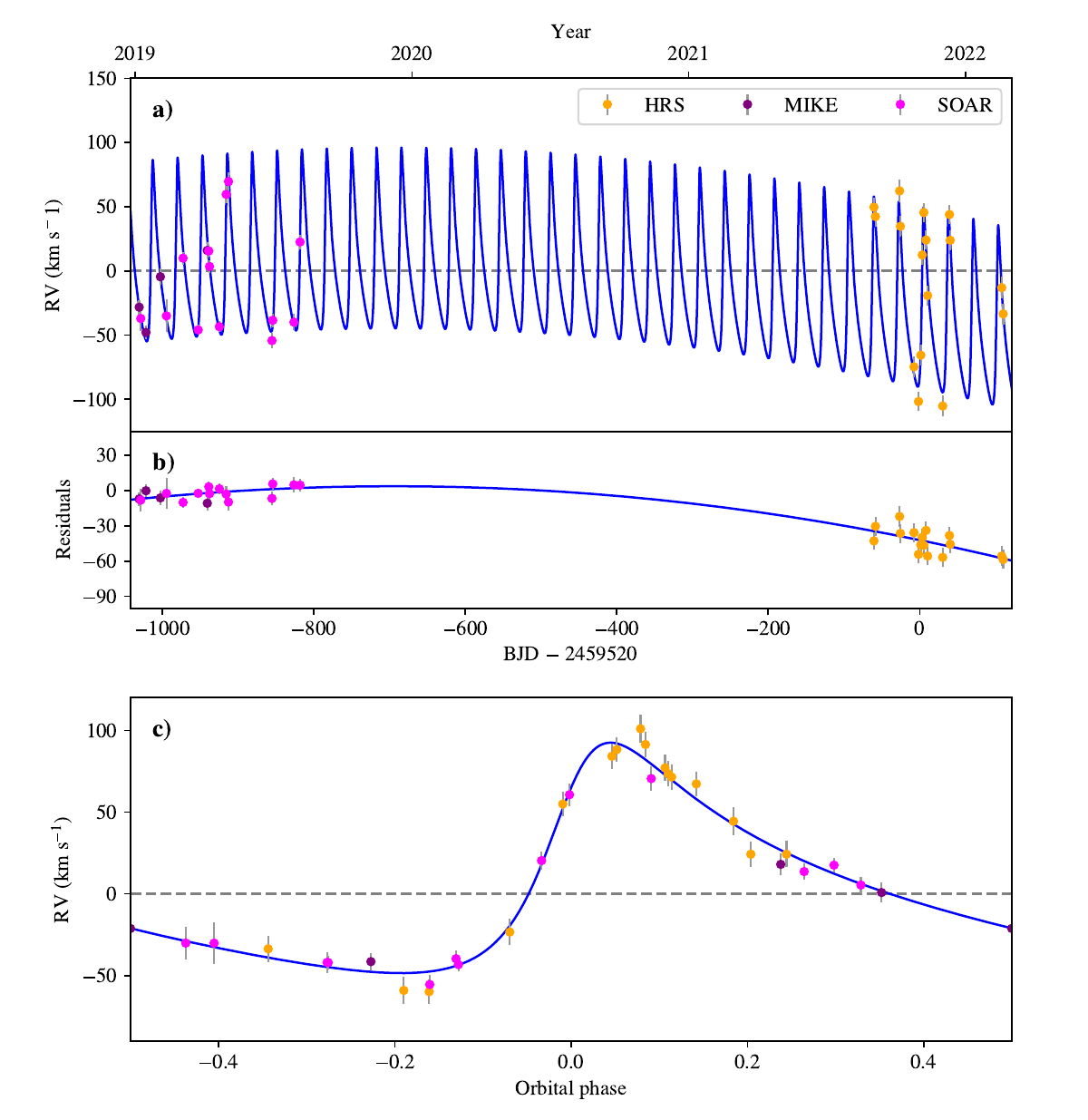}
   \caption{Optimized spectroscopic solution for the RVs of the primary component of ExtEV. The points on all panels are marked with different colors depending on the spectrograph (labeled in the top panel). The blue curve represents the best-fit model. (a) RV curve including orbital solution and the long-term changes. (b) RV curve with the orbital solution subtracted. (c) RVs phased with the orbital period given in Table~\ref{table:rv-fit} after freeing from the long-term changes of RVs shown in panel (b).}
   \label{fig:rv-solution}
\end{figure*}

\subsection{Spectroscopic orbit}\label{sect:rv-solution}
In order to obtain spectroscopic parameters of the primary's orbit, we used the RV data described in Sect.~\ref{sect:Time-series optical spectroscopy and radial velocities} together with those reported by J21.\footnote{The RVs were obtained by J21 using the Magellan Inamori Kyocera Echelle spectrograph \citep[MIKE,][]{2003SPIE.4841.1694B} and the Goodman spectrograph mounted at SOuthern Astrophysical Research telescope \citep[SOAR,][]{2004SPIE.5492..331C}.} The data were fitted with \texttt{RadVel} software \citep{2018PASP..130d4504F}, which uses a maximum a posteriori optimization method to find the best orbital solution and MCMC sampling to estimate errors. During the fitting, the orbital period, $P_{\rm orb}$, was treated as a fixed parameter, while the argument of periastron, $\omega$, the half-range of the RV changes of the primary component, $K_1$, and the time of the periastron passage, $T_{\rm peri}$, were treated as free parameters. Given that we use RVs obtained with three different spectrographs (SALT/HRS, MIKE, and SOAR), during the modeling procedure we allowed the corresponding systemic velocities ($\gamma_{\rm HRS}$, $\gamma_{\rm MIKE}$, and $\gamma_{\rm SOAR}$) to be independent. Furthermore, we noticed that the residuals from the initial fit revealed the presence of a long-term trend. We fitted this trend by adding a parabolic term that takes into account both the linear ($\dot{\gamma}$) and quadratic ($\ddot{\gamma}$) changes of the systemic velocity. Due to the significant dispersion of RVs, which is not due to measurement errors, but originates from the intrinsic variability of the primary component (TEOs and stochastic variability), we also added the so-called `jitter' term ($J_{\rm HRS}$, $J_{\rm MIKE}$ and $J_{\rm SOAR}$) as an additional parameter fitted separately for each dataset. This parameter accounts for the underestimated error associated with each RV measurement. As both $J_{\rm MIKE}$ and $J_{\rm SOAR}$ were consistent with zero within the errors\footnote{This is the most likely because J21 reported the errors of their RV measurements, which already accounted for $J_{\rm MIKE}$ and $J_{\rm SOAR}$ terms.}, we finally set their values to zero and fitted only $J_{\rm HRS}$. The results of the fit are given in Table~\ref{table:rv-fit} and shown in Fig.~\ref{fig:rv-solution}.

\begin{table}
\caption{Orbital spectroscopic solution for ExtEV based on RVs from SALT/HRS and the data presented by J21.}
\label{table:rv-fit}      
\centering                         
\begin{tabular}{l c}        
\hline\hline                 
\noalign{\smallskip}
Parameter & Optimized value with error\\
\noalign{\smallskip}
\hline    
\noalign{\medskip}
$P_{\rm orb}$\,(d) & 32.83016 (fixed) \\
\noalign{\smallskip}
$e$ & 0.513(28)\\
\noalign{\smallskip}
$\omega$\,($\degr$) & 308(4) \\
\noalign{\smallskip}
$K_1$\,(km\,s$^{-1}$) & 69.8(32) \\
\noalign{\smallskip}
$T_{\rm peri}$\,(BJD) & 2459523.75(21) \\
\noalign{\smallskip}
$\gamma_{\rm HRS}$\,(km\,s$^{-1}$) & 313(24) \\
\noalign{\smallskip}
$\gamma_{\rm MIKE}$\,(km\,s$^{-1}$) & 299(15) \\
\noalign{\smallskip}
$\gamma_{\rm SOAR}$\,(km\,s$^{-1}$) & 291(12) \\
\noalign{\smallskip}
$\dot{\gamma}$\,($10^{-3}$\,km\,s$^{-1}$\,d$^{-1}$) & $-$36(33) \\
\noalign{\smallskip}
$\ddot{\gamma}$\,($10^{-3}$\,km\,s$^{-1}$\,d$^{-2}$) & $-$0.09(4) \\
\noalign{\smallskip}
$J_{\rm HRS}$\,(km\,s$^{-1}$) & 9.9(28) \\
\noalign{\smallskip}
\hline
\noalign{\smallskip}
$f(M)$ (M$_\sun)^{\ast}$ & 0.74(5) \\
\noalign{\smallskip}
\hline                                   
\end{tabular}
\tablefoot{$^{\ast}$ Calculated from the values of $P_{\rm orb}$, $K_1$ and $e$.}
\end{table}

Based on the obtained solution, we calculated the mass function for this system, which amounts to $f(M)=0.74\pm0.05$\,M$_\sun$. This value is significantly larger than $0.51^{+0.07}_{-0.06}$\,M$_\sun$ given by J21. We also emphasize that the quadratic term is non-zero at the level of 2$\sigma$, implying that the ExtEV may be indeed a triple hierarchical system, with a tertiary component in a wide orbit with a relatively long orbital period. However, like the secondary component, the tertiary remains undetected in the ExtEV spectra.

\subsection{Geometry of the system during periastron passage}\label{sect:RLOF}
Theoretically, the large amplitude of light changes observed in ExtEV could be caused by the extreme non-linear tidal distortion of the primary component. This would mean a situation in which, when passing through the periastron, the primary component would be close to or filling its Roche lobe. This is exactly the scenario that was investigated by ML23 through hydrodynamic simulations. However, the number of assumptions made by these authors, for example, the lack of radiative losses, the adopted parameters of the system selected in the way that artificially increases tidal interaction, and the highly approximate method of transition from the hydrodynamic model to the light curve, do not allow this solution to be considered certain. Moreover, the authors claimed that the deep minimum in the light curve of ExtEV originates from a tidal bulge lifted almost directly towards the observer during the periastron passage. This conclusion was the result of a mistake that was made by using $\omega$ that was 180$\degr$ different from the true value. The mistake has been corrected in erratum \citep{ML23erratum}, but the updated figures reveal a smaller amplitude of light changes than originally announced by ML23 and the huge tidal bulge is no longer pointing toward the observer near the periastron.

ML23 adopted most of the system parameters from J21, including the rotation period of the primary component of about 4.4\,d. It turned out that only with an exceptionally fast rotation of the primary component, that is,  significantly larger than the pseudo-synchronous rotation rate, can Roche lobe overflow (RLOF) be observed in their simulations, followed by a subsequent non-linear collapse of a massive tidal wave generated by this close approach. However, KS22 (their Sect.~2.4) have already pointed out that there is no reasonable basis for the claim that the primary component of ExtEV rotates with a period of 4.4\,d. This value is the result of misclassifying one of the `red-noise' maxima in the frequency spectrum of the TESS data of ExtEV as the rotational frequency. This is evidenced by the lack of a significant maximum around frequency corresponding to the period of 4.4\,d in the frequency spectrum calculated by KS22 (their Figs.~4 and 10) using a much longer TESS light curve than that used by J21. We demonstrate in Sect.~\ref{sect:discussion-rotation} that the rotation period of the primary component amounts to at least about 8\,d. Next, the model presented by ML23 predicts that the frequency spectrum of the light curve should be dominated by the signal corresponding to the rotational frequency, which is not present in the TESS data. Finally, their hydrodynamic model is unable to reproduce one of the most characteristic features of the ExtEV light curve, that is, the TEO which is the 25th harmonic of the orbital frequency. All this makes the ML23 model questionable.

From an observational point of view, a very strong tidal distortion leading to the RLOF of the primary component near the periastron should induce a dramatic gravitational darkening \citep{1924MNRAS..84..665V,2018maeb.book.....P} of its outermost layers. Hence, one can expect heterochromatic effects in multi-color photometry, which should reveal the dependence of the amplitude of the ExtEV light curve on wavelength. Thanks to the LCOGT photometry (Fig.~\ref{fig:ExtEEV_LCO_2}), and the TESS photometry (Fig.~\ref{fig:tess_lc}), we can confidently conclude that the amplitude and shape of the `heartbeat' feature remains unchanged from Johnson $U$ to NIR TESS band at approximately 0.02\,mag level. This rather excludes the occurrence of strong temperature changes (of the order of a few $10^3$\,K) on the surface of the primary component. This conclusion is further supported by the lack of depth changes in the \ion{He}{i} lines with the orbital phase (Fig.~\ref{fig:OI_lines}), which we examined in our SALT/HRS spectra. The helium lines are very sensitive to changes in gas temperature \citep{2009ssc..book.....G}, so any strong tidal distortions on the surface of the primary component should translate into significant changes in the depths of \ion{He}{i} and \ion{He}{ii} lines. The depths of ionized metal lines (for example, the absorption lines of \ion{C}{iii}) do not change significantly in our spectra either. The phenomenon of light changes in ExtEV appears to be gray, inconsistent with the extreme tidal distortion and non-linear breaking of tidal waves proposed by ML23. We note, however, that for relatively small ellipsoidal distortion of the primary component during periastron passages, we do not expect to detect any heterochromatic effects in our multi-color light curves, as both the Johnson $U$ and TESS passbands lie on the Rayleigh-Jeans tail of its spectrum.

\begin{table}
\caption{Parameters and their distributions used to generate the histogram of $\vartheta$ parameter presented in Fig.~\ref{fig:RLdiff}.}
\label{table:RL-simulations}      
\centering                         
\begin{tabular}{l l}        
\hline\hline             
\noalign{\smallskip}
Parameter & Distribution\\
\noalign{\smallskip}
\hline 
\noalign{\medskip}
$M_1$/M$_\sun$&$\mathcal{U}(25,45)$ \\
\noalign{\smallskip}
$R_1$/R$_\sun$&$\mathcal{N}(30.0,2.2)$ \\
\noalign{\smallskip}
$f(M)$/M$_\sun$&$\mathcal{N}(0.74,0.05)$ \\
\noalign{\smallskip}
$e$&$\mathcal{N}(0.51,0.03)$ \\
\noalign{\smallskip}
$i$&$\mathcal{U}(30\degr,75\degr)$ \\
\noalign{\smallskip}
$\mathcal{F}_{\rm rot}$&$\mathcal{N}(1,0.25)$ \\
\noalign{\smallskip}
\hline                                   
\end{tabular}
\tablefoot{Symbols of statistical distributions are the same as in Table~\ref{table:mesa-distributions}.}
\end{table}

To check whether the primary component may undergo RLOF at periastron, we used the method presented by \cite{2007ApJ...660.1624S} to calculate its volume-equivalent Roche radius at periastron, $R_{\rm L}$\footnote{Calculation of this kind is purely dynamical. It does not account for the so-called Eddington factor \citep[e.g.][]{2011A&A...535A..56G}, which may slightly lower the values of $R_{\rm L}$ due to the radiation pressure-dominated atmosphere of a blue SG. We also neglect any velocity fields superimposed by TEOs.}. Unlike the \cite{1983ApJ...268..368E} formula, the method of \cite{2007ApJ...660.1624S} takes into account the non-zero eccentricity of the orbit. We use the quantity $\vartheta=(R_{\rm L}-R_1)/R_1$ as the RLOF indicator. A negative value of this quantity indicates RLOF at the periastron, while a positive value corresponds to the situation in which the system maintains a detached geometry throughout the entire orbital cycle.

\begin{figure}
   \centering
   \includegraphics[width=\hsize]{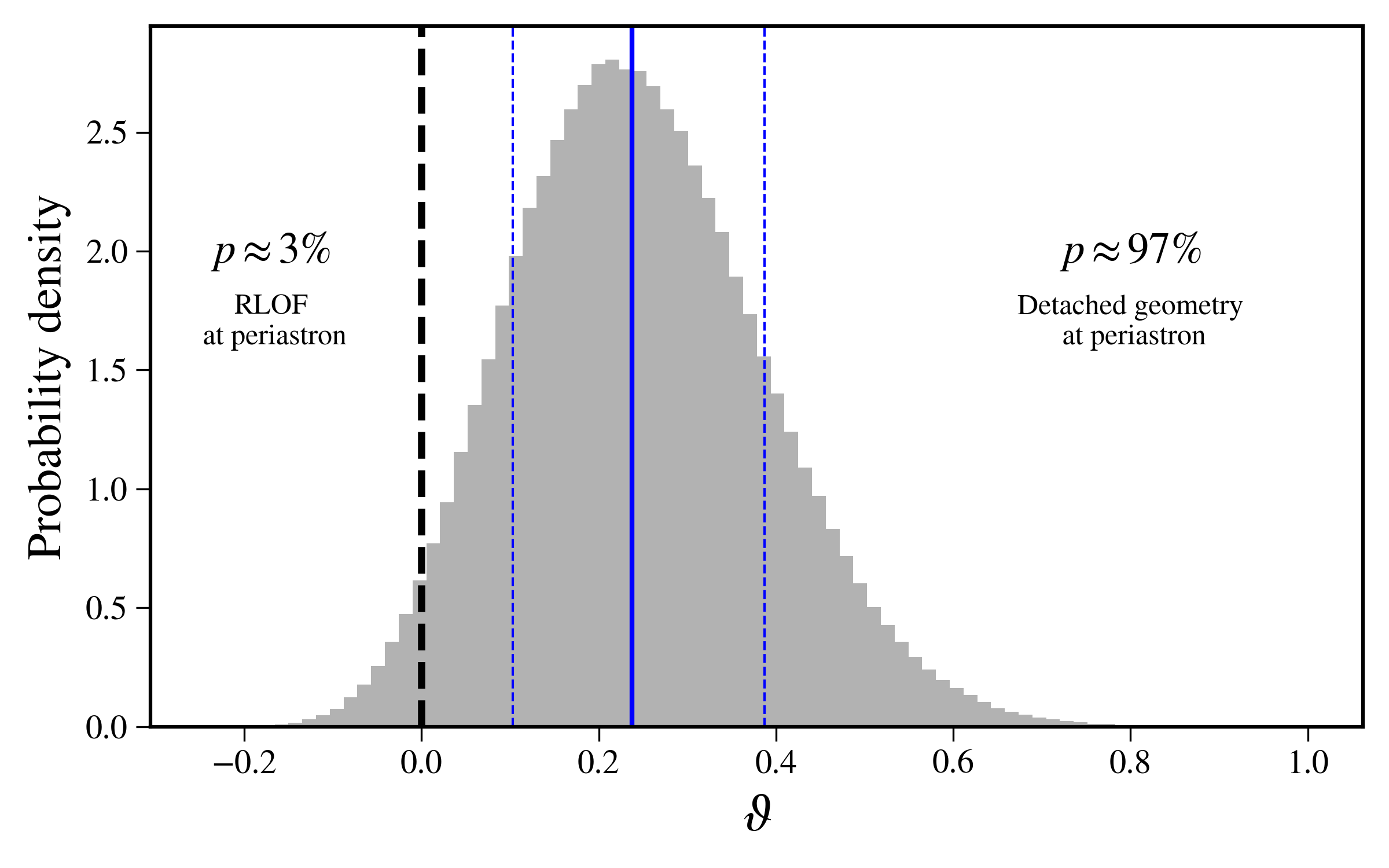}
   \caption{Histogram of the distribution of parameter $\vartheta=(R_{\rm L}-R_1)/R_1$, resulting from the simulations described in Sect.~\ref{sect:RLOF}. The black dashed vertical line indicates $\vartheta=0$, when the system becomes semi-detached. The blue vertical lines indicate quantiles at levels of 0.16, 0.5, and 0.84. Only about 3\% of samples correspond to an RLOF at periastron, while their majority ($\sim$97\%) point towards detached geometry at periastron.}
   \label{fig:RLdiff}
\end{figure}

Since we do not know precisely all the parameters of the ExtEV system, it is necessary to perform a statistical simulation to understand the distribution of the quantity $\vartheta$. The initial parameters of our simulation are listed in Table~\ref{table:RL-simulations}. Parameters such as $M_1$, $R_1$, $f(M)$, and $e$ and their ranges were adopted according to the results presented above. Since the inclination, $i$, is unknown, we assumed a realistic range of $30\degr$ to $75\degr$. Lower inclinations would correspond to unrealistically large masses of the primary component, while with higher values, we could expect photospheric eclipses, which are not observed. The last parameter in Table~\ref{table:RL-simulations}, $\mathcal{F}_{\rm rot}$, stands for the ratio of the rotational angular velocity of the primary component and its orbital angular velocity at periastron. The range adopted by us for $\mathcal{F}_{\rm rot}$ allows for rotation rates lower than pseudo-synchronous. The results are presented in the form of a histogram in Fig.~\ref{fig:RLdiff}. As can be seen, only $\sim$3\% of our models are characterized by RLOF at periastron. The majority of them ($\sim$97\%) remain detached throughout the entire range of orbital phases. The most probable value of $\vartheta$ amounts to approximately 0.23, suggesting that the primary component likely fills only about 50\% of its Roche lobe volume at periastron.

\subsection{No disk around the primary component}\label{sect:disk}
\begin{figure}
   \centering
   \includegraphics[width=0.89\hsize]{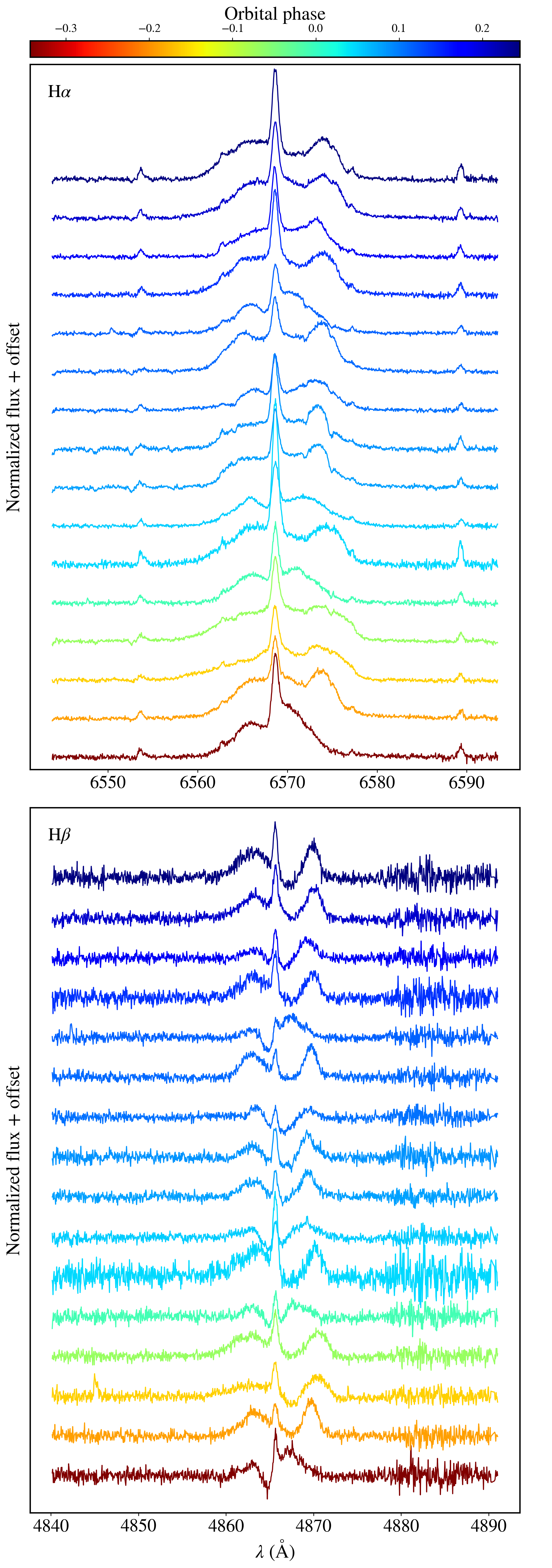}
   \caption{Fragments of SALT/HRS spectra of ExtEV near the H$\alpha$ (upper panel) and H$\beta$ (lower panel) lines. The spectra are color-coded according to the orbital phase (the color bar is located at the top of the figure), with zero phase corresponding to the periastron passage. The narrow emission in the center of the spectrum is of nebular origin.}
   \label{fig:Ha_Hb}
\end{figure}

The possibility of a periodically fading and re-emerging decretion disk in the ExtEV system was strongly suggested by J21 to explain the double-peaked Balmer emission lines. These are particularly visible in the H$\alpha$ and H$\beta$ lines (Fig.~\ref{fig:Ha_Hb}) in the form of broad, disjoint, and asymmetric wings accompanying the narrow emission peaks that appear in our spectra due to the nebular emission of the LMC. J21 concluded that such a disk must destabilize and disappear near the periastron passage, but reappear shortly thereafter, to generate such features throughout almost the entire orbital period. However, as can be easily seen in Fig.~\ref{fig:Ha_Hb} (lower panel), our spectra contradict this theory. There is no doubt that the emission in the H$\beta$ line persists even immediately after periastron passage and is not significantly weaker than just before phase zero.

To further investigate the plausibility of the scenario in which the emission features of the H$\alpha$ and H$\beta$ lines are attributed exclusively to the Keplerian disk around the primary component, we carried out simplified modeling in which we tried to estimate the actual location of the circumstellar disk and its geometry necessary to reproduce observable shapes of spectral lines. The aim of this experiment is not to determine the exact parameters of the disk, but to try to answer the question of whether it is crossed by the orbit of the secondary component, assuming that the two emitting components in the Balmer lines originate from the disk. We based our model on the following assumptions: (1) the disk forms a uniform, optically thin ring around the primary component, (2) it is aligned with the orbital plane, and (3) the rotation of the disk is Keplerian, so the velocity of the gaseous component at a distance $r_{\rm el}$ from the primary can be expressed as
\begin{equation}
    \label{apxeq:1}
    \centering
    v_\mathrm{el} = \sqrt{\frac{G M_1}{r_\mathrm{el}}},
\end{equation}
where $G$ is the universal gravitational constant. To obtain synthetic emission profiles, we associated each surface element of the disk with an identical, constant emitter of either H$\alpha$ or H$\beta$ photons with fixed wavelengths corrected for the radial velocity of the primary component. The resulting orbital velocity field -- uniquely defined by the width of the disk, $w_\mathrm{D}$, and the distance of its inner edge from the primary component, $d_\mathrm{D}$ -- was then projected into the radial direction and used to obtain the density distribution $\mathcal{H}\left(\lambda, d_\mathrm{D},w_\mathrm{D}\right)$ via relation: 
\begin{equation}
    \label{apxeq:2}
    \lambda_\mathrm{obs} \simeq \lambda_\mathrm{em}\left( 1 + \frac{v_\mathrm{rad}}{c} \right),
\end{equation}
which is a non-relativistic approximation of the Doppler formula. Here, $\lambda_\mathrm{em}$ is the emitted wavelength, $\lambda_\mathrm{obs}$ is the observed one, $v_\mathrm{rad}$ denotes the radial component of the element's velocity, and $c$ is the speed of light in vacuum. 

\begin{figure*}
   \centering
   \includegraphics[width=0.9\hsize]{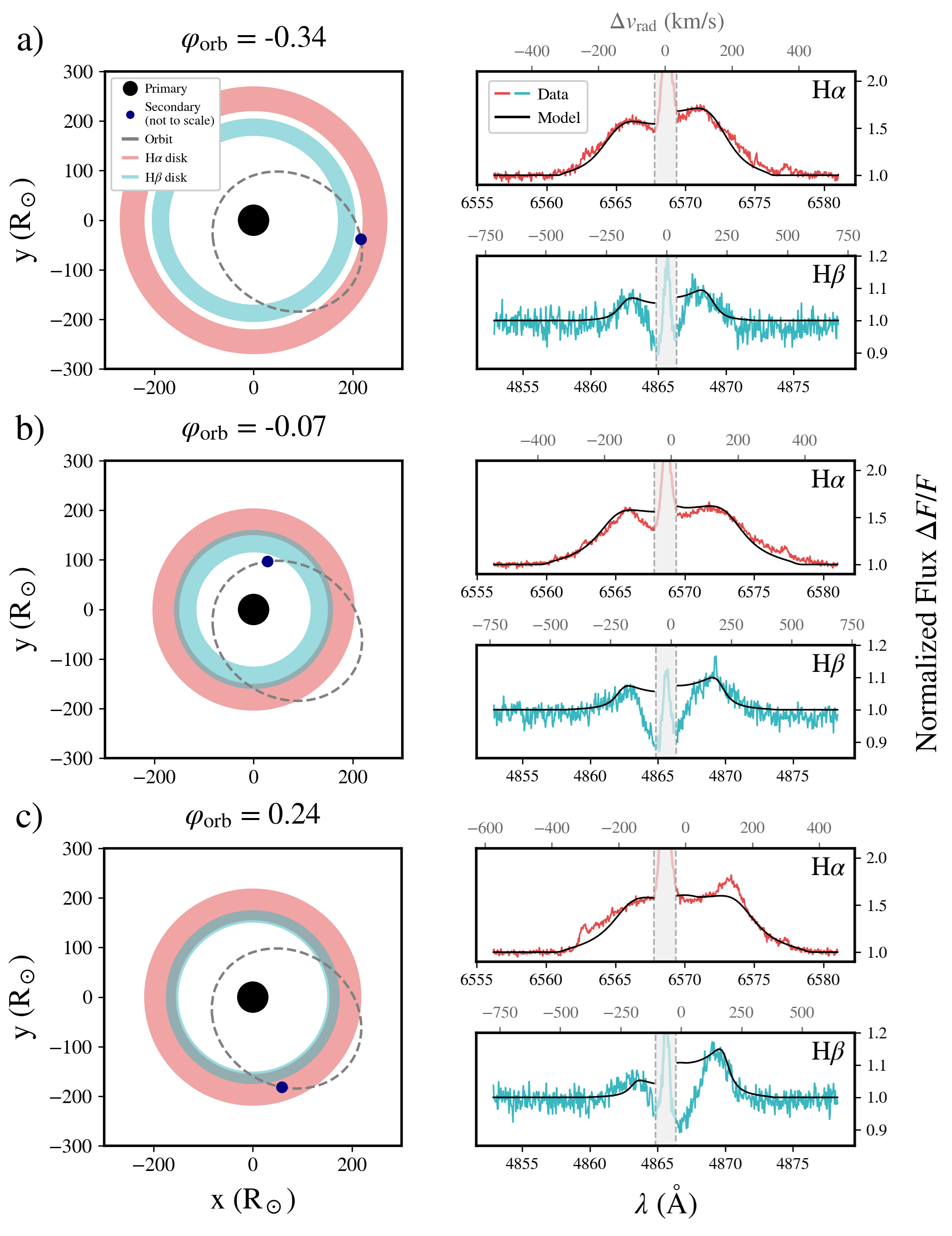}
   \caption{Models of the Keplerian rings around the primary component of ExtEV. Each set of figures (a, b, and c) corresponds to a different orbital phase $\varphi_{\rm orb}$ provided above each set. The right-hand panels show the observed H$\alpha$ (salmon) and H$\beta$ (aqua) lines to which the theoretical profiles (black curves) were fitted. For each spectrum, two rings were fitted separately, one responsible only for the H$\alpha$ emission and the other for the H$\beta$ emission. The narrow nebular emission of the LMC (gray region bounded by vertical dashed lines) was excluded from the fit. The left-hand panels show the resulting ring geometry compared to the size of the primary component (black circle) and the size of the orbit of the secondary component (dashed ellipse, secondary component size is not to scale). More details and discussion can be found in the main text.}
   \label{fig:disk}
\end{figure*}

To develop a model that could be directly fitted to the ExtEV spectra, the resulting probability distribution $\mathcal{H}\left(\lambda, d_\mathrm{D},w_\mathrm{D}\right)$ had to be divided into two separate blue- and red-shifted distributions, namely $\mathcal{H}_\mathrm{B}$ and $\mathcal{H}_\mathrm{R}$, and later convolved with the intrinsic emission line profile. For this purpose, we chose the Voigt function
\begin{equation}
    \label{apxeq:3}
    \mathcal{V}(\lambda,\sigma_G,\gamma_L) = \int\limits_{-\infty}^{\infty} \mathcal{G}(\lambda',\sigma_G)\, \mathcal{L} (\lambda - \lambda',\gamma_L) \,\mathrm{d} \lambda',
\end{equation}
where $\mathcal{G}$ and $\mathcal{L}$ denote the Gaussian and Lorentzian profiles, respectively. Their widths, parameterized by the standard deviation $\sigma_G$ and the scale factor $\gamma_L$, are used to account for thermal and pressure broadening, as well as any possible small-scale inhomogeneities or turbulent motions of the gas. The final model takes the following form
\begin{multline}
    \label{apxeq:4}
    \mathcal{M} = A_\mathrm{B} \,\mathcal{H}_\mathrm{B}(\lambda,d_\mathrm{D},w_\mathrm{D}) \ast \mathcal{V}(\lambda,\sigma_G,\gamma_L) \\ + A_\mathrm{R} \,\mathcal{H}_\mathrm{R}(\lambda,d_\mathrm{D},w_\mathrm{D}) \ast \mathcal{V}(\lambda,\sigma_G,\gamma_L),
\end{multline}
and is described by six parameters: $d_\mathrm{D}$, $w_\mathrm{D}$, $\sigma_G$, $\gamma_L$, and two additional scaling factors, $A_\mathrm{B}$ and $A_\mathrm{R}$, which take into account the heights of both peaks, separately the blue-shifted and red-shifted. We fitted this model to the SALT spectra using Monte Carlo sampling, based on a random walk in the parameter space with wide, uniform priors. We obtained the best-fitting profile by simply selecting the one that minimizes the $\chi^2$ discriminant. The results are shown in Fig.~\ref{fig:disk}, in which we compared three spectra at different orbital phases with fitted theoretical line profiles, along with visualizations of the sizes and locations of the corresponding rings. As can be seen, our experiment results in a disk that is intersected by the orbit of the secondary component. In such a configuration, it is difficult to imagine a disk that can form around the primary component and survive periodic and permanent gravitational perturbations from the secondary component. Even if such a disk were indeed not aligned with the orbital plane, the companion star should still intersect it, effectively destabilizing its structure. We therefore conclude that the emission features in ExtEV are unlikely to be due to the presence of a Keplerian disk around the primary. The observed shapes of the H$\alpha$ and H$\beta$ lines must originate from another source.

\section{Model of light variability}\label{sect:lc-model}
In the previous sections of this paper, we have argued that the non-linear model of tidal interactions presented by ML23 has serious shortcomings that question its ability to reliably explain the variability of ExtEV (Sect.~\ref{sect:RLOF}). We have also presented arguments against the presence of a disk around the primary component (Sects.~\ref{sect:Non-B[e] status of ExtEV} and \ref{sect:disk}). However, there are several indications of a significant mass loss in this system, which may be the direct cause of large-amplitude changes of the brightness observed in ExtEV. We therefore investigated an alternative model of the optical variability of ExtEV based on a high wind-driven mass-loss rate, enhanced by the presence of a nearby massive companion and its periodic periastron passages.

\begin{figure}
   \centering
   \includegraphics[width=\hsize]{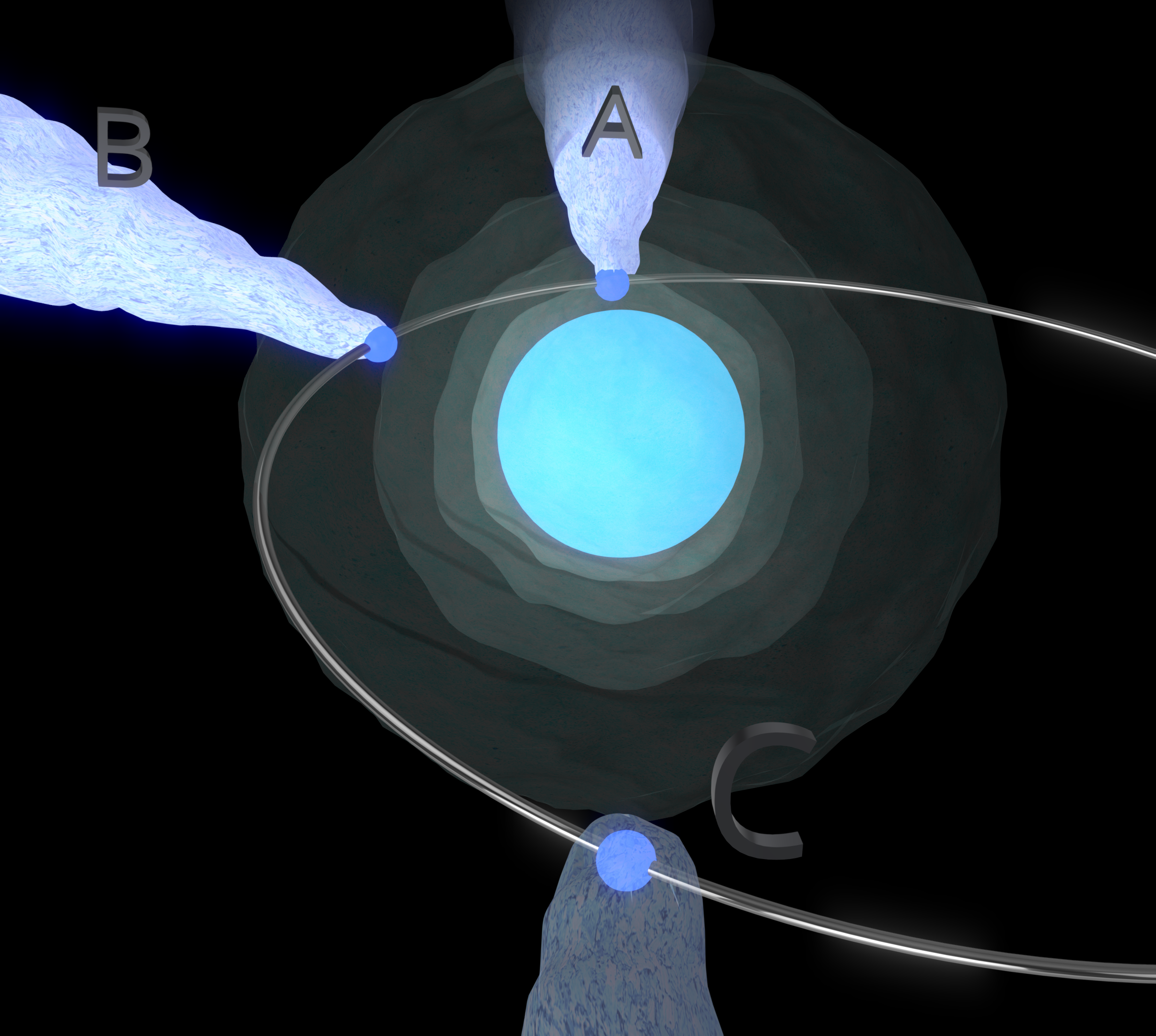}
   \caption[]{Artistic representation of the orbit (silver ellipse) of the secondary component of ExtEV (dark blue circle) around the primary component (light blue circle near the center). The size of the orbit and both components are to scale. The orientation of the orbit corresponds to the view of the ExtEV in the sky (i.e., as seen by the observer on Earth). The primary component is the source of a slow but dense stellar wind (semi-transparent spheres, concentric with the primary component), which, after colliding with the surface of the wind of the secondary component, forms a WWC cone, a turbulent structure, with its apex located near the secondary component. Three orbital phases are labeled, corresponding to: the superior conjunction (A), periastron passage (B), and inferior conjunction (C). More details can be found in the main text.}
   \label{fig:3d-rendering}
\end{figure}

We decided to check whether the variability of ExtEV can be explained by a model analogous to the one that explains the photometric variability of HD\,38282 system \citep{2021A&A...650A.147S}, which is characterized by $e=0.506$ and $\omega=304.6\degr$, very similar to those of the ExtEV's orbit. This system and ExtEV show surprisingly similar light curves. Figure~\ref{fig:3d-rendering} presents the dimensions and geometry of the ExtEV system to scale, along with three characteristic points in the secondary's orbit, labeled A (superior conjunction), B (periastron passage), and C (inferior conjunction). In our scenario, the decrease in brightness in the light curve of ExtEV has its origin in an atmospheric eclipse, when the light from the secondary component and the WWC cone is obscured by the dense stellar wind emitted by the blue SG (Fig.~\ref{fig:3d-rendering}, phase A). Both components are sources of stellar winds with completely different properties. As a B-type SG, the primary component emits an intense but relatively slow stellar wind that collides with the fast but low-density wind of the secondary component, typical for O-type dwarfs. As a result of the collision of the two winds, a geometrically thin WWC cone is formed very close to the photosphere of the secondary component. There is also a chance that the wind from the primary component is strong enough compared to the wind from the secondary that it will cause the front of the WWC cone, particularly at periastron, to collapse directly onto the surface of the secondary component, further eroding its photosphere. The gas in the WWC cone is so hot that it can be considered completely ionized, at least in the vicinity of the secondary component. The free electrons present in it can effectively scatter the light emitted from both components and direct it toward the observer. Additionally, hot gas flowing along the WWC cone eventually cools and recombines, emitting light in specific emission lines. Phase B in Fig.~\ref{fig:3d-rendering} corresponds to the situation when the WWC cone is the densest and hottest at the time of periastron passage. This is the source of the excess emission observed in ExtEV. After passing the periastron, in phase C, the secondary component is partially obscured by the dense and cooler parts the WWC cone. In fact, we expect the WWC cone to be significantly curved near periastron passage due to the relatively fast orbital motion of the secondary compared to the velocity of stellar wind.

\subsection{Analytical model}\label{sect:the analytical model}
The total brightness of ExtEV as a function of the true anomaly, $\nu$, can be approximated as follows:
\begin{equation}\label{eq:m}
    m\approx-2.5\log\left\{ F_1+\left[ F_2+F_{\rm WWC}(\nu) \right]e^{-\tau(\nu)} \right\}+C',
\end{equation}
where $F_1$, $F_2$, and $F_{\rm WWC}$ denote the flux from the primary component, from the secondary component, and from the WWC cone, respectively. The $\tau(\nu)$ denotes the optical depth in the wind of the primary component, integrated along the line of sight from the observer to the secondary component. $C'$ is the zero point of the magnitude scale. Equation~(\ref{eq:m}) assumes that the size of the secondary component is much smaller than that of the primary component. (According to our results, $R_1$ is about six times larger than $R_2$, which allows us to make this assumption.) Moreover, we also assumed that the brightest part of the WWC cone is in the close proximity of the secondary component. For this reason, the $F_{\rm WWC}$ in our model is also affected by the factor $e^{-\tau}$ because the excess flux emitted from the WWC cone can be obscured by the primary's wind. The relative change in the total brightness of ExtEV can be obtained from Eq.~(\ref{eq:m}) as
\begin{equation}\label{eq:delta m}
    \Delta m\approx-2.5\log\left\{ 1+\left[ (F_2/F_1)+(F_{\rm WWC}^{\rm peri}/F_1) \psi(\nu) \right]e^{-\tau(\nu)} \right\},
\end{equation}
where $F_{\rm WWC}^{\rm peri}$ is the excess emission of the WWC cone during periastron passage, while $\psi(\nu)$ stands for the function that modulates this excess emission over the orbital cycle and needs to be defined. Our model assumes isotropic and completely ionized wind of the primary component with a standard exponent of its $\beta$-velocity profile, $\beta=1$. According to these assumptions and theoretical formulae derived by \citet[][their Sect.~3]{1996AJ....112.2227L}, modified later by \citet[][their Sect.~3.5]{2021A&A...650A.147S} to account for elliptical orbits, it is possible to calculate $\tau(\nu)$ analytically from the following formula:
\begin{equation}\label{eq:tau}
\begin{aligned}
    \tau={} & \frac{(1+X_{\rm H})\sigma_{\rm e}\dot{M}_{\rm 1,wind}}{8\pi m_{\rm p}v_{1,\infty}r}\frac{2}{\sqrt{(1-\varepsilon^2)(1-b^2)}}\left[ \arctan\sqrt{\frac{1+b}{1-b}}+\right.\\
    & \left. \arctan\left( \sqrt{\frac{1+b}{1-b}} \tan \frac{\arcsin\varepsilon}{2} \right) \right],
\end{aligned}
\end{equation}
where $X_{\rm H}$ is the mass fraction of hydrogen in the wind of the primary component, $\sigma_{\rm e}$ is Thompson electron scattering cross-section, $m_{\rm p}$ is mass of the proton, $v_{1,\infty}$ denotes the terminal velocity of the primary's wind, and $r$ is the instantaneous distance between components. Parameters $\varepsilon$ and $b$ are defined as follows
\begin{equation}\label{eq:epsilon}
    \varepsilon=\sin i \cos (\nu + \omega + \pi/2),
\end{equation}
and
\begin{equation}\label{eq:b}
    b=\frac{R_1/r}{\sqrt{1-\varepsilon^2}}.
\end{equation}
Distance $r$ between the components can be easily obtained according to the well-known formula for elliptical orbits,
\begin{equation}\label{eq:r}
    r=\frac{a(1-e^2)}{1+e\cos \nu},
\end{equation}
where the semi-major axis of the relative orbit, $a$, has to be calculated from Kepler's third law. The last necessary ingredient, i.e. $\psi(\nu)$, can be approximated in the following way. As mentioned above, the WWC cone shines primarily due to the scattering of the primary component's light on free electrons, whose flux varies approximately as $r^{-2}$. Furthermore, the density of WWC cone decreases approximately as $r^{-1}$ \citep[e.g.][]{1996ApJ...469..729C,2022ApJ...933....5I}. Therefore, we can expect that these two processes (i.e., the decrease in gas density in the WWC cone and the decrease in the number of photons reaching it from the primary component with increasing $r$) will largely determine the changes in the brightness of the WWC cone. Hence, we can reasonably assume that \citep[after][]{2021A&A...650A.147S}
\begin{equation}\label{eq:psi}
    \psi(\nu)\approx \left( \frac{r_{\rm peri}}{r} \right)^{\gamma_{\rm WWC}} =  \left( \frac{1+e\cos \nu}{1+e} \right)^{\gamma_{\rm WWC}},
\end{equation}
where $r_{\rm peri}$ is the periastron distance between the components and we expect an exponent $\gamma_{\rm WWC}\gtrsim3$, for the reasons described above.

The reason why we cannot directly use the formulae for $\Delta m$ from \cite{1996AJ....112.2227L} is that these authors assumed optically thin wind (i.e.~$\tau\ll 1$), which probably is not the case for the intense wind of ExtEV. Moreover, our Eq.~(\ref{eq:delta m}) intentionally does not contain the term associated with the atmospheric eclipse of the primary component by the wind emanating from the secondary. In our preliminary analysis, we noticed that even after including this additional atmospheric eclipse, the impact of the secondary's wind on the light curve of ExtEV is negligible.

\subsection{Optimized solution}\label{sect:fit of our model}
\begin{figure*}
   \centering
   \includegraphics[width=\hsize]{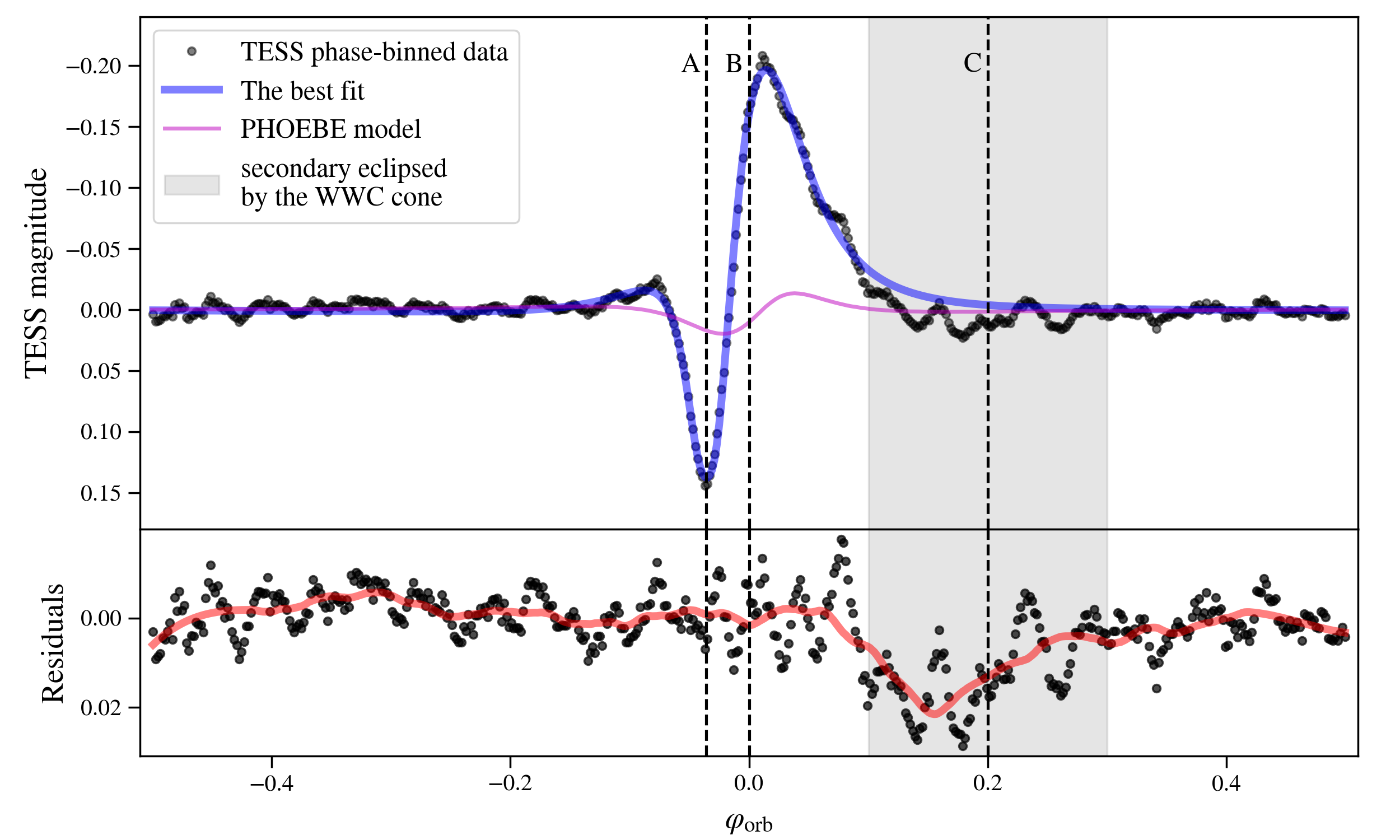}
   \caption{Plot summarizing the fit of our analytical variability model (Sects.~\ref{sect:the analytical model} and \ref{sect:fit of our model}) to the TESS light curve of ExtEV. The upper panel shows the TESS light curve binned in the orbital phase (black dots) with the best-fit model (blue curve) superimposed. The pink line corresponds to the variability model generated with PHOEBE. Phases A, B, and C from Fig.~\ref{fig:3d-rendering} are shown with vertical dashed lines and labeled. The vertical shaded region marks the range of orbital phases where we expect the secondary component to be partially obscured by the WWC cone. The lower panel shows the residuals from the best fit and their smoothed version (red curve). The zero phase corresponds to the time of periastron passage. More details can be found in the main text.}
   \label{fig:tess_best_fit}
\end{figure*}

To fit the analytical variability model described in the previous section, we phased the TESS light curve (Sect.~\ref{sect:time-series photometry}) with the orbital period and binned it in 0.002 intervals to preserve the original shape of the heartbeat-like feature as accurately as possible (Fig.~\ref{fig:tess_best_fit}). Due to the presence of TEOs with different amplitudes (KS22), part of their signal remains visible in the binned light curve, while the stochastic variability has been effectively averaged out.

\begin{table}
\caption{Optimized parameters of our analytical model of the light curve.}
\label{table:lc-fit}      
\centering                         
\begin{tabular}{l r}        
\hline\hline                 
\noalign{\smallskip}
Parameter & Optimized value with error\\
\noalign{\smallskip}
\hline     
\noalign{\medskip}
$i$\,($\degr$) & $65.93^{+0.38}_{-0.40}$\\
\noalign{\smallskip}
$\log[({\rm d}M/{\rm d}r)_\infty$/(M$_\sun$\,R$_\sun^{-1})]$ & $-9.004^{+0.009}_{-0.009}$\\
\noalign{\smallskip}
$F_2/F_1$ & $0.153^{+0.008}_{-0.006}$\\
\noalign{\smallskip}
$F_{\rm WWC}^{\rm peri}/F_1$ & $0.506^{+0.007}_{-0.007}$\\
\noalign{\smallskip}
$\gamma_{\rm WWC}$ & $4.76^{+0.05}_{-0.05}$\\
\noalign{\smallskip}
$\Delta \varphi_{\rm orb}$\,($10^{-2}$) & $-0.702^{+0.016}_{-0.016}$\\
\noalign{\smallskip}
$C$\,($10^{-3}$) & $0.02^{+0.33}_{-0.33}$\\
\noalign{\smallskip}
$\log\xi$ & $-2.284^{+0.016}_{-0.016}$\\
\noalign{\smallskip}
\hline                                   
\end{tabular}
\end{table}

For practical reasons, we fitted the $\Delta m'$ function to such binned light curve, which slightly differed from the form indicated in Eq.~(\ref{eq:delta m}). The fitted function was transformed from $\Delta m$ to the following form
\begin{equation}\label{delta m fit}
    \Delta m' = \Delta m(\varphi_{\rm orb}+\Delta\varphi_{\rm orb}) - {\rm median}\left\{\Delta m(\varphi_{\rm orb}+\Delta\varphi_{\rm orb})\right\} + C,
\end{equation}
where $\Delta \varphi_{\rm orb}$ denotes a small phase shift between the TESS data and the model due to the finite precision of our $T_{\rm peri}$, and $C$ is a constant that accounts for any subtle vertical shift. Since our model does not account for the `eclipse' of the secondary component by the WWC cone near the inferior conjunction\footnote{Accounting for this phenomenon with a simple analytical description seems difficult to achieve, as its key aspects include the exact shape and orientation of the WWC cone and its density distribution. These cannot be easily obtained without sophisticated hydrodynamic modeling, which includes radiative processes.}, we excluded the phase range between 0.1 and 0.3 from the fitting procedure. However, we would like to emphasize that, after several tests, this did not significantly affect the fitting results.

Before starting the fitting process, the calculation of $\tau$ (Eq.~(\ref{eq:tau})) required two additional steps. First, we had to select a specific value of $X_{\rm H}$. In our case, we have set the canonical value $X_{\rm H}=0.7$. Secondly, $\dot{M}_{\rm 1,wind}$ and $v_{1,\infty}$ are highly correlated, so it is not possible to fit these two parameters independently. We can only optimize their ratio, which corresponds to the mass of the wind contained in a shell of thickness ${\rm d}r$ at a distance from the primary component where the wind reaches its terminal velocity. We denote this ratio as $({\rm d}M/{\rm d}r)_\infty$. We performed fitting and estimated its errors using a method analogous to that described in Sect.~\ref{sect:sed-fitting}. This time, however, we used a different form of the nuisance parameter. We assumed that the error associated with each point in the binned light curve is equal to $\xi$. In this way, we have incorporated in the fitting the scatter due to TEOs, which was not originally included in our model. The free parameters for the optimization were therefore $i$, $\log({\rm d}M/{\rm d}r)_\infty$, $F_2/F_1$, $F_{\rm WWC}^{\rm peri}/F_1$, $\gamma_{\rm WWC}$, $\Delta \varphi_{\rm orb}$, $C$, and $\log\xi$. We emphasize that parameters such as $e$, $\omega$ and $f(M)$, taken from Table~\ref{table:rv-fit}, were fixed during fitting, strongly constraining possible solutions. We also recall that $M_1$ was set to 35\,M$_\sun$, but discuss the impact of changing this value in Appendix~\ref{appendix:different M1}. The results of our fitting are shown in Fig.~\ref{fig:tess_best_fit}, while the optimized parameters with their errors are presented in Table~\ref{table:lc-fit}. We include a corner plot from the corresponding MCMC analysis of the solution in Appendix~\ref{appendix:corner plot}.

\subsection{Interpretation of the solution}
It is surprising that the simple analytical model we presented above fits the TESS light curve (Fig.~\ref{fig:tess_best_fit}) so well. This is evidenced by the residuals, which, due to the partially remaining TEOs, oscillate around zero, except in the vicinity of the inferior conjunction, which we write about in more detail below. Although the model has five free parameters that determine its shape\footnote{Parameters such as $\Delta \varphi_{\rm orb}$, $C$ and $\log\xi$ do not directly shape the morphology of the synthetic light curve.}, we want to emphasize that the fit has been obtained without modifying the orbital parameters. With the exception of the orbital inclination, the others were fixed during optimization. To determine to what extent phenomena such as ellipsoidal distortion and irradiation effect may contribute to the ExtEV's light curve, we generated a synthetic TESS light curve using PHOEBE\,2.4 software (pink curve in Fig.~\ref{fig:tess_best_fit}). For this purpose, we used the following set of input parameters. The $P_{\rm orb}$, $e$ and $\omega$ were taken from Table~\ref{table:rv-fit}, while $i$ was taken from Table~\ref{table:lc-fit}. For the primary component we adopted $M_1=35$\,M$_\sun$ and $R_1$ from Table~\ref{table:sed-fit}, while for the secondary component we took $M_2=13$\,M$_\sun$ and $R_2=5$\,R$_\sun$. The effective temperatures of the primary and secondary components were taken from J21. The atmospheres of both components were treated as black bodies. The gravity-darkening coefficients, as well as the bolometric albedos, were set to 1 for both components. The limb-darkening coefficients were assumed, according to the tables from \cite{2017A&A...600A..30C} for the TESS passband. We assumed that both components rotate pseudo-synchronously. As can easily be seen in Fig.~\ref{fig:tess_best_fit}, proximity effects make a secondary contribution to the high amplitude of the light variability of ExtEV.

Although we originally fitted a value of $\log({\rm d}M/{\rm d}r)_\infty$, one can easily estimate the resulting value of $\dot{M}_{\rm 1,wind}$ from this parameter, realistically assuming $v_{1,\infty}\approx 1000$\,km\,s$^{-1}$ \citep[e.g.][]{2000ARA&A..38..613K}. We obtained that the primary component loses mass due to the stellar wind at a rate of $\sim4.5\cdot10^{-5}$\,M$_\sun\,{\rm yr}^{-1}$. Such a high mass loss rate makes the primary component of ExtEV unique and explains many of its observed properties. We provide further discussion on this topic in Sect.~\ref{sect:discussion-mdot}.

Since the atmospheric eclipse is very sensitive to the inclination of the orbit, this parameter is determined with relatively high precision (Fig.~\ref{fig:cornerplot}). Therefore, we can expect $M_2\approx13$\,M$_\sun$. We have verified that with this orbital configuration and the radius of the secondary component $R_2\approx5$\,R$_\sun$, the system is on the borderline between a pure atmospheric eclipse and an atmospheric eclipse combined with a so-called grazing photospheric eclipse. The obtained $F_2/F_1$ ratio of about 0.15 theoretically indicates that the spectra of the components can be disentangled. However, due to, for example, fast rotation of the secondary component, its lines may still be strongly diluted in the composite spectrum.

The residuals (Fig.~\ref{fig:tess_best_fit}, phase C) reveal a decrease in brightness of about 0.02 mag near inferior conjunction, which is not accounted for in our model. We interpret this decrease as attenuation of light from the secondary component by turbulent structures in the WWC cone that are present in this range of orbital phases between the observer and the secondary component.

\subsection{Scatter of the residuals over the orbital cycle}\label{sect:scatter of residuals}

\begin{figure*}
   \centering
   \includegraphics[width=\hsize]{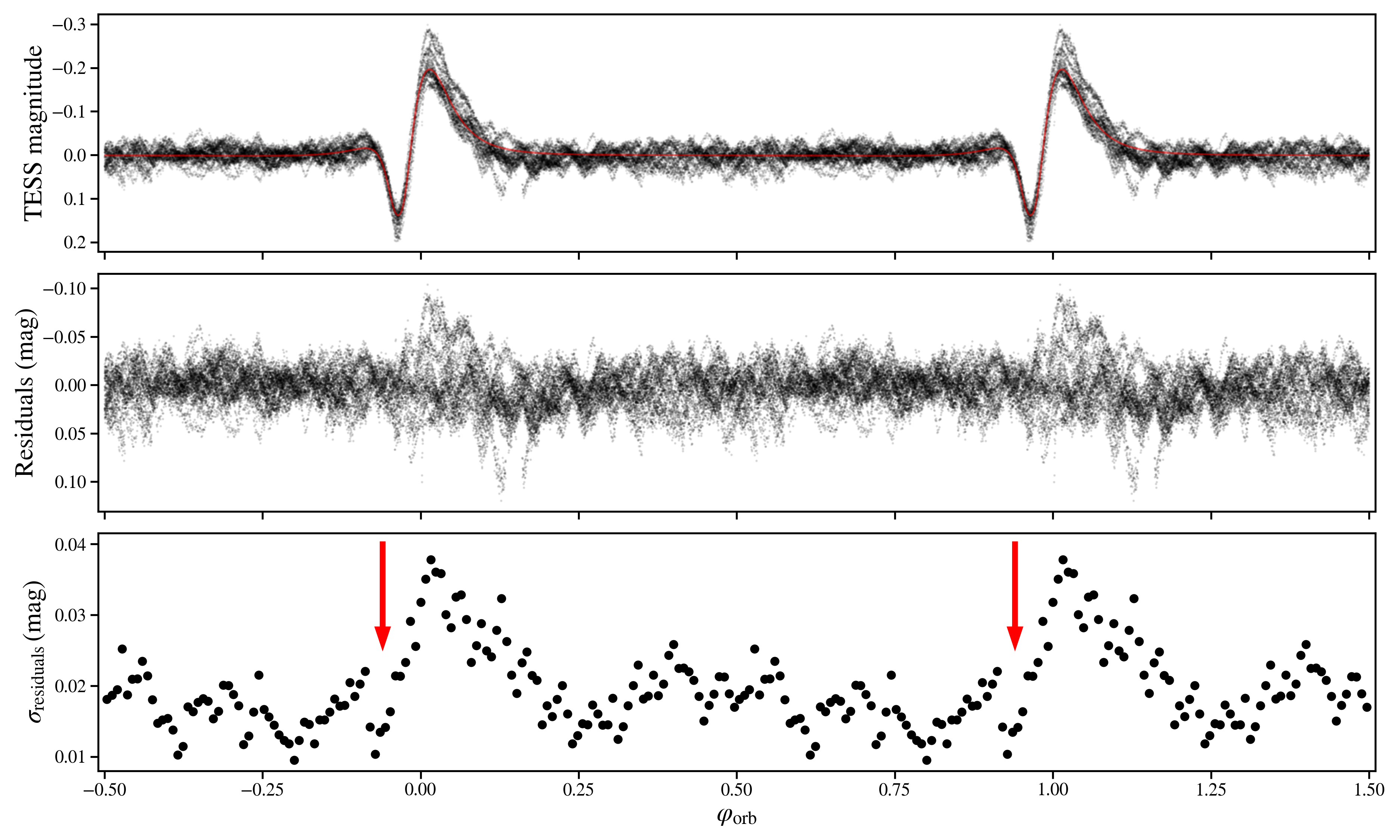}
   \caption{Changes of the scatter of residuals from the fit of our model to the light curve of ExtEV. The upper panel shows the TESS light curve (black points) folded with the orbital period, while the red curve corresponds to the best-fitting model from Fig.~\ref{fig:tess_best_fit}. Phase $\varphi_{\rm orb}=0$ corresponds to the periastron passage. The middle panel presents the residuals from the fit shown in the upper panel. The lower panel shows the changes in the local scatter of the residuals, calculated as the standard deviation of the residuals in 0.007 phase intervals. The red arrows indicate a sudden decrease in the scatter of residuals around the phase of superior conjunction.}
   \label{fig:std_in_phase}
\end{figure*}

With the fitted variability model for ExtEV, we can subtract it from the original TESS light curve and examine the behavior of the residuals by calculating the local scatter as the standard deviation of the residuals over a specified range of orbital phases. Figure~\ref{fig:std_in_phase} shows the result of this procedure. It is clear that the scatter of the residuals (Fig.~\ref{fig:std_in_phase}, bottom panel) is not random but depends on the orbital phase. This scatter is due to at least three phenomena: the TEOs (as they change their amplitude with time), the stochastic variability of the primary component \citep[caused, for example, by internal gravity waves,][]{2013ApJ...772...21R}, and the turbulent gas structures in the WWC cone. In Fig.~\ref{fig:std_in_phase} we can see that the scatter of the residuals is the largest during the periastron passage. We interpret this increase as reflecting increased turbulence of the gas in the WWC cone, causing the light curve to `flicker' with greater intensity. This seems to be a plausible explanation for several reasons. Firstly, it is at the periastron passage when the secondary component moves through the densest medium at the highest relative velocity. Secondly, the WWC cone has the highest density at this particular orbital phase, so it effectively scatters light. Finally, the WWC cone in the periastron has the smallest geometric thickness and can be subject to the so-called thin-shell instability, making the gas motion highly turbulent \citep[e.g.][]{2021Galax..10....4K}.

A very interesting feature of the scatter of the residuals in Fig.~\ref{fig:std_in_phase}, indicated by the red arrows in the lower panel, is the clear and rapid decrease in the scatter around the phase of superior conjunction ($\varphi_{\rm orb}\approx-0.04$). This decrease can be easily interpreted based on the model we have adopted. When the secondary component with the WWC region is behind the primary component, there is a partial occultation of the WWC cone by the atmosphere, and possibly also by the photosphere, of the primary component. Since it is mainly the WWC cone that is responsible for the changes in the scatter of the residuals, the increasing contribution of the turbulent WWC emission in this phase to the total brightness of the ExtEV system suddenly decreases, leading to a significant reduction of scatter in the residuals.

\section{Discussion}\label{sect:discussion}

\subsection{Wind mass-loss rate}\label{sect:discussion-mdot}
The mass-loss rate we derived for ExtEV, $4.5\cdot 10^{-5}$\,M$_\sun\,{\rm yr}^{-1}$, is exceptionally high compared to a typical blue SG with parameters similar to the primary component of ExtEV. To see how this compares with theoretical predictions, we calculated the values of mass-loss rate based on three selected wind mass-loss prescriptions and the physical parameters of the primary component that we determined in our study. The result of this comparison is presented in Table~\ref{table:comparison of mdots}. It is clear that the wind generated by the primary component can be up to two orders of magnitude more intense than what contemporary theory predicts for a single star of this type. If our proposed scenario of light variability is correct, this implies that the primary component of ExtEV exhibits a wind mass loss rate typical of WR stars or red SGs from the tip of the asymptotic giant branch\footnote{Typically, the mass-loss rate for WR stars is of the order of $10^{-5}$\,M$_\sun\,{\rm yr}^{-1}$ \citep{2020MNRAS.499..873S}, while for some of the AGB stars it can reach even a $10^{-4}$\,M$_\sun\,{\rm yr}^{-1}$ level \citep{2018A&ARv..26....1H}.} (AGB), although it is still relatively close to the MS and its spectra definitely do not exhibit spectral features of WR stars.

\begin{table}
\caption{Comparison between the observed and predicted wind mass-loss rates of the primary component.}
\label{table:comparison of mdots}       
\centering                         
\begin{tabular}{c l}        
\hline\hline                 
\noalign{\smallskip}
$\log(\dot{M}_{\rm 1,wind}/($M$_\sun\,{\rm yr}^{-1}))$ & Reference\\
\noalign{\smallskip}
\hline
\noalign{\smallskip}
$-$4.35 & This work\\
\noalign{\smallskip}
$-$5.63 & \text{\cite{2001A&A...369..574V}}\\
$-$6.13 & \text{\cite{2021A&A...647A..28K}}\\
$-$6.27 & \text{\cite{2023A&A...676A.109B}}\\
\noalign{\smallskip}
\hline                                   
\end{tabular}
\end{table}

A significant mass-loss rate found in the ExtEV should result in noticeable changes in the orbital period. Indeed, KS22 have shown that the orbital period of the ExtEV is decreasing at a rate of approximately 11\,s\,yr$^{-1}$. They also estimated that if we attribute the observed changes in orbital period solely to mass loss due to the stellar wind of the primary component\footnote{The estimation performed by KS22 (their Sect.~5.3) assumed that the wind intensity varies significantly over the orbital phase and is highest at periastron due to the proximity of the companion. Therefore, the authors adopted a model of perfectly non-conservative mass loss in the form of periodic `ejection' of matter from the primary component only at the periastron passage.}, the rate should be of the order of $10^{-4}$\,M$_\sun\,{\rm yr}^{-1}$. Given that tides and TEOs also contribute to the shortening of the orbital period, the aforementioned value of $\dot{M}_{\rm 1,wind}$ fits very well with the proposed high mass loss in the ExtEV. We conclude that ExtEV is a WWC binary system with a deep atmospheric eclipse and strong excess emission during periastron passage.

In this context, an intriguing question arises regarding the source of such a high mass loss by the primary component. We can suspect that this is due to the presence of a close and massive companion. Because of the significant eccentricity of its orbit, along with the change in orbital phase, the primary's Roche lobe periodically approaches the surface of the star, which can significantly enhance its stellar wind \cite[e.g.][]{1988MNRAS.231..823T,2014PASJ...66...82L,2020ApJ...902...85M}. Moreover, the primary component shows the presence of relatively high-amplitude TEOs, which may also be responsible for the enhancement of the stellar wind. Thus, it appears that ExtEV is an extremely rare massive binary system that offers us a unique opportunity to observe a short phase of its evolution, during which eccentricity, proximity to a massive companion, rotation and the presence of TEOs significantly enhance the stellar wind, and their mutual interaction determines the evolution of the whole system.

\subsection{Rotation rate of the primary component}\label{sect:discussion-rotation}
Knowing the inclination angle of the orbit of ExtEV and assuming that the rotational axis of the primary component is perpendicular to the orbital plane, we can derive linear equatorial velocity $v_{\rm eq}\approx190\pm38\,$km\,s$^{-1}$, bearing in mind that the observed $v_{\rm eq}\sin i=174\pm34\,$km\,s$^{-1}$ (J21). This is rather an upper limit, since J21 measured $v_{\rm eq}\sin i$ from a single line of \ion{He}{i}\,$\lambda\,4922\,\AA$, neglecting its distortion and broadening caused by several TEOs. The value of $v_{\rm eq}$ implies that the primary component is likely to rotate with a period of about $8.0\pm2.0$\,d, provided that it is not significantly oblate due to rotation. This is twice as long as the rotation period adopted by ML23 to obtain the large amplitude of light variations in their hydrodynamic model.

Knowing $v_{\rm eq}$ and the other parameters of ExtEV obtained in our paper, we can calculate the critical rotation period of the primary component following \cite{2019A&A...625A..88G}
\begin{equation}\label{eq:omega_crit}
    P_{\rm crit}=2\pi\left[\frac{GM_1}{R_{\rm eq}^3}\left(1-\Gamma^{3/2}\right)\right]^{-1/2},
\end{equation}
where $R_{\rm eq}$ is the equatorial radius of the primary component if it would rotate critically. We can realistically estimate $R_{\rm eq}\approx1.5R_1$ \citep[e.g.][]{1959cbs..book.....K}, assuming the Roche model of a rotating star and that its polar radius, $R_{\rm p}$, does not change with rotation rate, i.e. $R_{\rm p}\approx R_1$. In turn, $\Gamma=\kappa L_1/(4\pi cGM_1)$ stands for the so-called Eddington factor. In the definition of $\Gamma$, $\kappa$ is the flux-weighted mass absorption coefficient in the photosphere of the primary component. Since the temperature of the sub-photospheric layers increases rapidly, we can assume that $\kappa$ is dominated by electron scattering in the outer layers, i.e. $\kappa\approx\kappa_{\rm e}\approx 0.34\,{\rm cm}^2\,{\rm g}^{-1}$. Thus, $\Gamma\approx0.5$ and we are left with $P_{\rm crit}\approx 9.2\pm1.2\,$d\footnote{In our calculations of the formal error of $P_{\rm crit}$ we assumed that standard deviation of $M_1=35\,$M$_\sun$ is of the order of $5\,$M$_\sun$, in line with the results presented in Sect.~\ref{sect:mass-estimation}.}. Theoretically, this value would indicate a critical rotation of the primary component. However, the problem is that if the primary component rotates (nearly) critically, our transformation of $v_{\rm eq}$ to $P_{\rm rot}$ using the relation $P_{\rm rot}=2\pi R_{\rm eq}/v_{\rm eq}$, where we assumed $R_{\rm eq}=R_1$, is no longer valid. To examine the impact of nearly critical rotation on the $P_{\rm rot}$ and $P_{\rm crit}$ presented above, we performed the following estimations. The value of $R_1$ obtained in Sect.~\ref{sect:sed-fitting} is a proxy for $L_1$, since we considered the effective temperature of the primary component to be known. The luminosity of a rotating primary component can be approximated as $L_1\approx4\pi R_{\rm p}^2\sigma_{\rm SB}T_{\rm p}^4\eta(\Omega_{\rm rot}/\Omega_{\rm crit})$, where $\eta(1)\approx 0.64$\footnote{According to the lecture notes on stellar rotation by J. P. Harrington, \url{https://www.astro.umd.edu/~jph/Stellar_Rotation.pdf}.} is the dimensionless factor accounting for the latitude-dependent surface temperature distribution, $\sigma_{\rm SB}$ is the Stefan-Boltzmann constant, and $T_{\rm p}$ stands for surface temperature at the poles. If the rotation of the primary component approaches critical rotation, the $R_1$ obtained by us in terms of SED fitting changes its original meaning to $R_1\sim\sqrt{\eta(1)}R_{\rm p}\approx\sqrt{0.64}R_{\rm eq}/1.5$, where we again applied the relation $R_{\rm eq}=1.5R_{\rm p}$. Finally, it leads us to $R_{\rm eq}\approx 1.87R_1$. This result translates into $P_{\rm rot}=14.9\pm3.7\,$d and $P_{\rm crit}=12.7\pm1.7\,$d, which points towards subcritical rotation, contrary to the previous result. We therefore conclude that without detailed modeling, it is difficult to unambiguously answer the question about the value of $\Omega_{\rm rot}/\Omega_{\rm crit}$.

We also checked how the rotation period of the primary component compares to the pseudo-synchronous rotation period, $P_{\rm ps}$, that can be expected under effective tidal action. Taking the value of $e$ from Table~\ref{table:rv-fit} and using the formulae derived by \cite{1981A&A....99..126H}, we obtained $P_{\rm ps}\approx 8.45\,$d. This value is satisfactorily consistent with the rotation period of the primary component we obtained under the assumption of a subcritical rotation rate.

\subsection{Remarks on X-ray signal and optical emission features}\label{sect:discussion-X-rays}
Wind-wind collision binaries are also known to emit significant X-ray flux, which can vary with the orbital phase \citep[see][for a comprehensive review on this topic]{2016AdSpR..58..761R}. The primary source of X-ray radiation in these systems is the front of the WWC cone, the temperature of which (depending on the properties of the system) may reach approximately $10^7$\,K \citep[e.g.][]{2011A&A...530A.119P}. Therefore, one would expect that since ExtEV is such an extreme case of a WWC binary, it would have a high X-ray luminosity, $L_X$, which should be the strongest at the periastron passage. Unfortunately, J21, after carefully examining the Swift X-ray Telescope data (their Sect.~3.6), did not find any statistically significant X-ray emission at the position of ExtEV. However, the authors derived an upper limit to the X-ray brightness, $L_X\lesssim 2.5\cdot10^{34}\,$erg\,s$^{-1}$. Comparing this value to other WWC binaries \citep[][and references therein]{2000Ap&SS.274..159C}, it turns out that it is difficult to detect even a strong X-ray signal with Swift due to the distance separating us from the LMC. ExtEV may therefore be characterized by significant X-ray emission, but an X-ray observatory with better capabilities than Swift is needed to detect it. Moreover, the extreme conditions in the WWC cone do not automatically result in the strong X-ray flux observed outside the system. This can happen for at least three reasons. First, dispersed gas and dust around ExtEV, but also the gas within the WWC cone, can effectively absorb X-ray photons generated by the collision of winds as is in the case of HD\,38282 \citep{2021A&A...650A.147S}. Secondly, if the front of the WWC cone falls near the periastron (or even throughout the orbital phase) onto the surface of the secondary component, there is a sharp drop in $L_X$ due to the lack of the hottest and densest region of the WWC, as is the case with $\eta$\,Car, which is also a WWC binary with a very high orbital eccentricity \citep{2012ApJ...746...73T}. Finally, as recently shown by \cite{2023MNRAS.526.3099M}, inverse Compton scattering occurring in the WWC shocked region, often neglected during the modeling of WWC binaries, can effectively suppress the X-ray emission. This is because the mentioned physical process acts as an additional cooling mechanism that softens the spectrum emerging from the densest parts of the WWC cone.

Although the optical spectra of the ExtEV exhibit strong emissions in the Balmer H$\alpha$ and H$\beta$ lines, we do not observe their counterparts in the helium or ionized metal lines, which are typical tracers of recombination regions in the WWC cone. However, in some \ion{He}{i} lines we do observe clear emission features located in the wings. As with X-rays, this does not necessarily imply that ExtEV is not a source of strong emissions in the aforementioned lines. As noted by \cite{2022RMxAA..58..403K} for the HD\,5980 WWC system, it is very difficult to determine whether the moderate wing emission is the result of two separate Doppler-shifted emission components or whether it is a strong and broad emission line, which is superimposed with almost equally strong absorption with a slightly narrower profile. Most likely, photons emitted in lines located in the WWC recombination zone encounter the dense, optically thick wind of the primary component, potentially leading to their effective absorption. For the reasons outlined here, the analysis of emission features in the ExtEV optical spectrum cannot be unambiguous without an advanced hydrodynamic model of the entire system, taking into account its dynamics, the structure of the WWC cone, stellar wind and radiative transfer. This issue is obviously beyond the scope of our paper and requires further examination using sophisticated, time-consuming simulations.

\section{Summary and conclusions}\label{sect:summary-and-conclusions}
We analyzed photometric properties, light curves from TESS and LCOGT, and SALT/HRS optical spectra for the massive eccentric binary system, MACHO\,80.7443.1718 (ExtEV), located in the LMC. This system shows extremely large brightness variations (Fig.~\ref{fig:tess_lc}), which raises numerous questions about their nature (Sect.~\ref{sect:introduction}). Our main findings can be summarized as follows.
\begin{enumerate}
    \item Thanks to the LCGOT light curves of ExtEV collected in the $U$, $B$, and $V$ bands (Sect.~\ref{sect:time-series photometry}, Fig.~\ref{fig:ExtEEV_LCO_2}), we conclude that the range of brightness changes and the shape of the light curve do not depend on wavelength over a broad range from NUV to NIR.
    \item SED of ExtEV reveals significant IR excess (Sect.~\ref{sect:sed-fitting}), especially prominent in the MIR bands (Fig.~\ref{fig:double-dust-sed-fit_v3}). Although the IR properties of ExtEV indicate an enhanced mass loss rate in the system, they do not match the properties of B[e] SGs (Sect.~\ref{sect:Non-B[e] status of ExtEV}, Fig.~\ref{fig:ir-photometry}). Moreover, in our SALT/HRS spectra we did not detect any emission lines, such as [\ion{O}{i}] lines, typical for the B[e] phenomenon  (Fig.~\ref{fig:OI_lines}).
    \item SED modeling resulted in the radius of the primary component $R_1\approx30$\,R$_\sun$, with a luminosity of about $\log(L_1$/L$_\sun)\approx5.82$ (Table~\ref{table:sed-fit}). The SED indicates also for the presence of an additional attenuation around $2760\,\AA$ of unknown origin (Fig.~\ref{fig:double-extinction_v3}).
    \item Determination of the initial and current mass of the primary component is difficult, primarily due to the strong influence of rotation and mass loss on its evolution (Sect.~\ref{sect:mass-estimation}). However, our simulations showed that the mass of the primary component at ZAMS most likely falls within the range of 27 to 55\,M$_\sun$. The present mass of the star lies between 25 and 45\,M$_\sun$ (Fig.~\ref{fig:HRD}). The evolutionary status of the primary component, as deduced from the comparison with evolutionary tracks,  corresponds to either hydrogen or helium core burning.
    \item We derived spectroscopic parameters of the primary's orbit combining archival RVs of J21 and RVs obtained from spectra collected with SALT/HRS (Sect.~\ref{sect:rv-solution}, Fig.~\ref{fig:rv-solution}, Table~\ref{table:rv-fit}). In particular, we derived a more precise value of the mass function $f(M)\approx0.74\pm0.05$\,M$_\sun$. We also found significant changes in the $\gamma$ velocity of ExtEV, an indication that the system is triple.
    \item We demonstrated that the assumption that the emission features in the H$\alpha$ and H$\beta$ lines originate in a Keplerian disk around the primary component leads to the location of the disk that would strongly interact with the secondary's orbit (Sect.~\ref{sect:disk}, Fig.~\ref{fig:disk}). We also argued that the primary component likely retains a detached geometry during the periastron passage (Sect.~\ref{sect:RLOF}, Fig.~\ref{fig:RLdiff}), and its rotation period of about 8\,d is consistent with pseudo-synchronous rotation ($P_{\rm ps}\approx 8.45$\,d; Sect.~\ref{sect:discussion-rotation}).
    \item The light curve of ExtEV can be successfully explained by a superposition of an atmospheric eclipse of the secondary component by the intense stellar wind of the primary and excess emission originating from scattered light on relatively dense structures of the WWC cone (Sect.~\ref{sect:lc-model}, Fig.~\ref{fig:tess_best_fit}). As a result of our modeling, we precisely determined the inclination of the ExtEV's orbit ($i\approx66\degr$) and the wind mass-loss rate from the primary component, $\log(\dot{M}_{\rm 1,wind}/($M$_\sun\,{\rm yr}^{-1}))= -4.35$ (Table~\ref{table:lc-fit}).
\end{enumerate}

Based on these results, we can draw the following conclusions.
\begin{enumerate}
    \item ExtEV is not a B[e] SG because its several key properties are significantly different from those of B[e] SGs. This also means that the extreme changes in the brightness of ExtEV cannot be explained by a disk surrounding the primary component.
    \item It seems unlikely that the light variability in ExtEV is caused by an extreme tidal distortion and subsequent non-linear breaking of tidal waves on the surface of the primary component, as proposed by ML23. First, the system most likely does not undergo RLOF at periastron, and the rotation rate of the primary component is at least twice lower than that assumed in the ML23 models. Second, the shape and peak-to-peak amplitude of the light curve of ExtEV are virtually the same in Johnson $UBV$ and TESS passbands, which rules out the presence of extreme tidal distortions of the primary component that would be accompanied by a significant gravitational darkening and hence heterochromatic effects in the light curve.
    \item Ellipsoidal distortion appears to be a secondary factor in shaping the light curve of ExtEV, and therefore the star is not an extreme case of an EEV system. In other words, it is not an extreme heartbeat star. According to our study, ExtEV may be in fact a unique WWC binary system in which a considerable mass loss by the primary component causes a remarkably large range of light variability close to periastron passage. Therefore, a particular care must be taken when classifying massive binary systems as EEVs based on their light curves, as some of them may actually be WWC systems mimicking EEVs.
    \item The rate of mass loss in the primary component that we obtained is two orders of magnitude greater than what the theory predicts. We can therefore suspect that the wind of the blue SG in ExtEV is tidally and rotationally enhanced, possibly with some additional interaction with TEOs. In this scenario, ExtEV is an extremely rare observational example of a massive binary system during a short but dramatic stage in its evolution, when the proximity of a massive companion leads to a sudden stripping of the primary's envelope without the need for RLOF. If this scenario holds in ExtEV, it is likely that other massive binary systems also experience such episodes of significant mass loss, even on the MS, without having to become WR stars or red giants. Thus, ExtEV can serve as an excellent laboratory for studying the mechanisms of wind enhancement in massive stars and predicting the impact of this enhancement on their evolution.
\end{enumerate}

In order to verify the variability scenario proposed in our study, we are going to perform IR interferometric observations of ExtEV and collect its UV spectra, which will help us to determine the parameters of its stellar wind, including an independent assessment of its mass-loss rate. In a final solution of the puzzling nature of these stars, it would also be beneficial to perform polarimetric measurements of this system in different orbital phases.

\begin{acknowledgements}
We would like to thank the anonymous referees for many useful comments and suggestions that helped to improve the publication. PKS would like to express his gratitude to Milena Ratajczak and Marcin Wrona for their assistance in preparing observing time proposals for ExtEV. PKS is also grateful to Morgan MacLeod for the fruitful discussion about the nature of ExtEV, and to Damien Gagnier for the discussion on some properties of rotating stars. Some of the observations reported in this paper were obtained with the Southern African Large Telescope (SALT) under program 2021-1-SCI-022 (PI: PKS). Polish participation in SALT is funded by grant No. MEiN nr 2021/WK/01. This work makes use of observations from the Las Cumbres Observatory global telescope network. PKS was supported by the Polish National Science Center grant no.~2019/35/N/ST9/03805. AP, PKS, and P\L{} would like to appreciate the financial support from the Polish National Science Center grant no.~2022/45/B/ST9/03862. TR was partly founded from budgetary funds for science in 2018-2022 in a research project under the program ,,Diamentowy Grant'', no. DI2018 024648. This publication makes use of VOSA, developed under the Spanish Virtual Observatory (\url{https://svo.cab.inta-csic.es}) project funded by MCIN/AEI/10.13039/501100011033/ through grant PID2020-112949GB-I00. VOSA has been partially updated by using funding from the European Union's Horizon 2020 Research and Innovation Programme, under Grant Agreement no.\,776403 (EXOPLANETS-A). The authors made use of the Strasbourg Astronomical Data Center (CDS) portal and the Barbara A.~Mikulski Archive for Space Telescopes (MAST) portal. This paper includes data collected by the TESS mission, which are publicly available from the MAST. This research has made use of ,,Aladin sky atlas'' developed at CDS, Strasbourg Observatory, France. This research has made use of the VizieR catalog access tool, CDS, Strasbourg, France. This research made use of NumPy \citep{harris2020array}, SciPy \citep{2020SciPy-NMeth}, Matplotlib \citep{Hunter:2007} and AstroPy \citep{2022ApJ...935..167A}.
\end{acknowledgements}

\bibliographystyle{aa}
\bibliography{bib}

\begin{appendix}

\section{The impact of change in $M_1$}\label{appendix:different M1}
Since the present mass of the primary component, $M_1$, is not known with high precision (Sect.~\ref{sect:mass-estimation}), we investigate how the change in $M_1$ affects the results presented in Sect.~\ref{sect:fit of our model}, especially the parameters presented in Table \ref{table:lc-fit} obtained assuming $M_1=35$\,M$_\odot$. For this purpose, we considered two other cases,  $M_1=30$\,M$_\sun$ and 40\,M$_\sun$. The fitting procedure remained unchanged. The effect of this experiment is summarized in Table~\ref{table:change of M1}. All parameters but inclination change only slightly with the change of $M_1$. The change of $i$ between the two solutions is about 1.9$\degr$ and is at the 5$\sigma$ level of the uncertainty of $i$. Therefore, we can safely state that the conclusions drawn in our study are virtually independent of the assumption of the value of $M_1$ in the realistic range of this parameter.

\begin{table}
\caption{Impact of the change of $M_1$ on the parameters of the fit described in Sect.~\ref{sect:fit of our model}.}
\label{table:change of M1}       
\centering                         
\begin{tabular}{l r r}        
\hline\hline                 
\noalign{\smallskip}
\multirow{2}{*}{Parameter} & \multicolumn{2}{c}{Optimized value with error}\\
\noalign{\smallskip}
&$M_1=30$\,M$_\sun$&$M_1=40$\,M$_\sun$\\
\noalign{\smallskip}
\hline
\noalign{\medskip}
$i$\,($\degr$) & $64.91^{+0.38}_{-0.39}$ & $66.77^{+0.36}_{-0.40}$\\
\noalign{\smallskip}
$\log[({\rm d}M/{\rm d}r)_\infty/($M$_\sun$\,R$_\sun^{-1})]$ & $-9.028^{+0.009}_{-0.009}$ & $-8.983^{+0.009}_{-0.009}$\\
\noalign{\smallskip}
$F_2/F_1$ & $0.155^{+0.008}_{-0.007}$ & $0.153^{+0.007}_{-0.006}$\\
\noalign{\smallskip}
$F_{\rm WWC}^{\rm peri}/F_1$ & $0.511^{+0.007}_{-0.006}$ & $0.505^{+0.007}_{-0.006}$\\
\noalign{\smallskip}
$\gamma_{\rm WWC}$ & $4.77^{+0.05}_{-0.05}$ & $4.75^{+0.05}_{-0.05}$\\
\noalign{\smallskip}
$\Delta \varphi_{\rm orb}$\,($10^{-2}$) & $-0.720^{+0.016}_{-0.015}$ & $-0.688^{+0.016}_{-0.015}$\\
\noalign{\smallskip}
$C$\,($10^{-3}$) & $0.02^{+0.33}_{-0.33}$ & $0.02^{+0.34}_{-0.33}$\\
\noalign{\smallskip}
$\log\xi$ & $-2.284^{+0.016}_{-0.015}$ & $-2.284^{+0.016}_{-0.016}$\\
\noalign{\smallskip}
\hline                                   
\end{tabular}
\end{table}

\section{Corner plot}\label{appendix:corner plot}
\begin{figure*}
   \centering
   \includegraphics[width=\hsize]{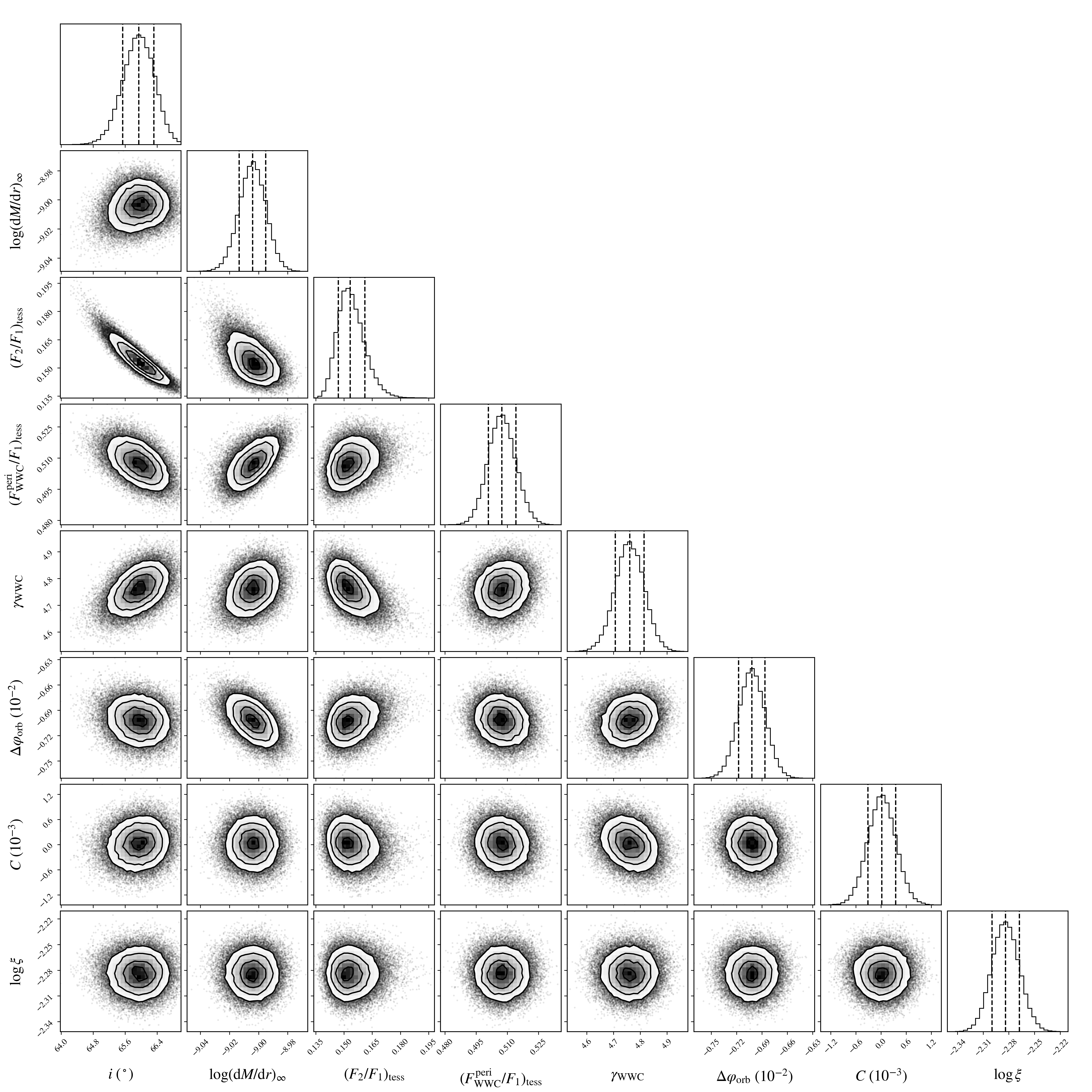}
   \caption{Corner plot resulting from our MCMC simulations described in Sect.~\ref{sect:fit of our model}. A series of vertical dashed lines in each histogram marks the position of 0.18, 0.5, and 0.84 quantiles.}
   \label{fig:cornerplot}
\end{figure*}

To generate the corner plot presented in Fig.~\ref{fig:cornerplot} we used 32 walkers, each with 100\,000 steps, from which 500 were considered as `burn-in' steps. Before generating the corner plot, each of the chains was `thinned' by half of the autocorrelation time, which amounted to about 100 steps. The solution does not reveal any significant correlations between the free parameters of the model, except for the obvious one between $F_2/F_1$ and $i$.

\end{appendix}

\end{document}